\documentstyle[12pt]{article}

\begin{document}
\hfuzz=10pt
\ifx\oldzeta\undefined
  \let\oldzeta=\zeta
  \def\zzeta{{\raise 2pt\hbox{$\oldzeta$}}}
  \let\zeta=\zzeta
\fi

\ifx\oldchi\undefined
  \let\oldchi=\chi
  \def\cchi{{\raise 2pt\hbox{$\oldchi$}}}
  \let\chi=\cchi
\fi

\ifx\oldxi\undefined
  \let\oldxi=\xi
  \def\xxi{{\raise 2pt\hbox{$\oldxi$}}}
  \let\xi=\xxi
\fi
 
\font\sstwelve=cmss12
\font\ssnine=cmss9
\font\sseight=cmss8
\newcommand\zZtwelve{\hbox{\sstwelve Z\hskip -4.5pt Z}}
\newcommand\zZnine{\hbox{\ssnine Z\hskip -3.9pt Z}}
\newcommand\zZeight{\hbox{\sseight Z\hskip -3.7pt Z}}
\newcommand\zZ{\mathchoice{\zZten}{\zZten}{\zZeight}{\zZeight}}
\newcommand\ZZ{\mathchoice{\zZtwelve}{\zZtwelve}{\zZnine}{\zZeight}}
 
\mathchardef\sigma="711B
\mathchardef\tau="711C
\mathchardef\nabla="7272

\font\twelvemb=cmmib10 scaled \magstep1
\font\tenmb=cmmib10
\font\ninemb=cmmib9
\font\sevenmb=cmmib7
\font\sixmb=cmmib6
\font\fivemb=cmmib5
\font\twelvesyb=cmbsy10 scaled \magstep1
\font\tensyb=cmbsy10
\font\dezb=cmmib10
\textfont9=\tenmb
\scriptfont9=\sevenmb
\scriptscriptfont9=\fivemb
\textfont10=\tensyb

\def\bn{{\fam10\nabla}}

\def\bm{\fam9}
 
\newcommand{\fracsm}[2]{{\textstyle\frac{#1}{#2}}}
\newcommand{\ur}[1]{(\ref{#1})}
\newcommand{\eq}[1]{eq.~(\ref{#1})}
\newcommand{\eqs}[2]{eqs.~(\ref{#1},~\ref{#2})}
\newcommand{\eqsss}[2]{eqs.~(\ref{#1}--\ref{#2})}
\newcommand{\Eq}[1]{Eq.~(\ref{#1})}
\newcommand{\Eqs}[2]{Eqs.~(\ref{#1},~\ref{#2})}
\newcommand{\Eqsss}[2]{Eqs.~(\ref{#1}--\ref{#2})}
\newcommand{\fig}[1]{Fig.~\ref{#1}}
\newcommand{\beq}{\begin{equation}}
\newcommand{\eeq}{\end{equation}}
\newcommand{\e}{\varepsilon}
\newcommand{\ee}{\epsilon}
\newcommand{\la}[1]{\label{#1}}
\newcommand{\SU}{$SU(2)~$}
\newcommand{\doublet}[3]{\: \left(\begin{array}{c} #1 \\#2
\end{array} \right)_{#3}}
\newcommand{\matr}[4]{\left(\begin{array}{cc}
#1 &#2 \\
#3 &#4                       \end{array} \right)}
\newcommand{\Pmax}{P_{max}}

\newcommand{\op}[1]{{\bf \hat{#1}}}
\newcommand{\opr}[1]{{\rm \hat{#1}}}
\newcommand{\At}{\tilde A}
\newcommand{\Bt}{\tilde B}
\newcommand{\Ct}{\tilde C}
 
\newcommand{\bra}[1]{\langle#1\vert}
\newcommand{\ket}[1]{\vert#1\rangle}
 
\newcommand{\C}[3]{C^{#1}_{#2,#3}}
\newcommand{\Sj}[6]{
\left\{\begin{array}{ccc}
 #1 & #2  & #3 \\
 #4 & #5  & #6
\end{array}\right\}}

\newcommand{\Nj}[9]{\left\{
\begin{array}{ccc}
 #1 & #2  & #3 \\
 #4 & #5  & #6 \\
 #7 & #8  & #9
\end{array}\right\}}
 
\newcommand{\Tj}[6]{
\left(\begin{array}{ccc}
 #1 & #2  & #3 \\
 #4 & #5  & #6
\end{array}\right)}

\newcommand{\vtr}[3]{\left(\begin{array}{c}
 #1\\
 #2\\
 #3
\end{array}\right)}

\newcommand{\irre}[3]
  {\langle #1\parallel #2\parallel#3\rangle}
 
\def\appendix{\par
 \setcounter{section}{0}
\def\thesection{Appendix}}

\thispagestyle{empty}
\begin{flushright}
RUB-TPII-62/93 \\
January 1994
\end{flushright}
\vspace{20 pt}
\begin{center}
{\Large\bf Fermion Sea Along the Sphaleron Barrier\\}
\vspace{40 pt}
{\large\bf
Dmitri Diakonov$^*$\footnote{Alexander von Humboldt
Forschungspreistr\"{a}ger
\vskip 5 pt
\noindent e-mail: \\
diakonov@lnpi.spb.su \\
maxpol@lnpi.spb.su   \\
peters@proton.tp2.ruhr-uni-bochum.de  \\
joergs@proton.tp2.ruhr-uni-bochum.de  \\
goeke@hadron.tp2.ruhr-uni-bochum.de},
Maxim Polyakov$^*$, \\
Peter Sieber$^\diamond$, J\"org Schaldach$^\diamond$
and Klaus Goeke$^\diamond$}
\end{center}
 
\vspace{20 pt}
\noindent
{\small\it $^*$St.~Petersburg
Nuclear Physics Institute, Gatchina, St.Petersburg 188350, Russia \\
$^\diamond$Inst. f\"ur Theor. Physik II, Ruhr-Universit\"at Bochum,
D-44780 Bochum, Germany}
\vspace{40 pt}
\abstract{We study the fermion Dirac sea along the minimal
energy path between topologically distinct vacua in the electroweak
theory. We follow in detail the interplay between the bound state and
the continuum in the fermion crea\-tion/\-an\-ni\-hi\-la\-tion, as
one passes from
one vacuum to another. We calculate the quantum correction to the
classical energy of the sphaleron barrier, arising from the fermion sea,
and its contribution to the baryon number violation rate for
non-zero temperatures. We find that the fermion fluctuations suppress 
that rate if the mass of the top quark is large enough.}
\newpage
 
\section{Introduction}
\setcounter{equation}{0}
\def\theequation{\arabic{section}.\arabic{equation}}

The fact that the potential energy of the Yang--Mills fields
is periodic in
a certain topological functional of the gauge fields, called the
Chern--Simons number
$N_{CS}$, was shown quite a long time ago by Faddeev \cite{F} and Jackiw and
Rebbi \cite{JR}. In an unbroken Yang--Mills theory (like QCD) the classical
energy barrier between the topologically distinct vacua can be made
infinitely small due to the scale invariance. In a spontaneously broken
theory (like the electroweak one) the scale invariance is explicitly
violated by the Higgs v.e.v., and the height of the barrier is of the
order of $m_W/\alpha$, where $m_W$ is the $W$-boson mass and
$\alpha=g^2/4\pi$ is the \SU gauge coupling.

Transitions from one vacuum to a topologically distinct one over
this barrier cause a change in the baryon and lepton number by one
unit per fermion doublet due to the axial anomaly \cite{tH}. Hence,
this transition is a baryon and lepton number violating process and
therefore of great physical significance. Although it is
suppressed under ordinary conditions \cite{tH}, its rate 
can become large at
high densities \cite{R,DP}, high temperatures \cite{AM} or high particle 
energies \cite{DiPol,DiPet}.
Thus, it could have occurred in the early universe \cite{Shap1,Shap2}, 
and an exact determination
of the transition rate is important to understand the excess of matter
over antimatter.

The static configuration of the Yang--Mills (YM) and Higgs ($H$) field
corresponding to the top of the barrier and having $N_{CS}=1/2$,
called sphaleron, was first found by Dashen, Hasslacher and Neveu \cite
{DHN} and, in the context of the electroweak theory, rediscovered by
Klinkhammer and Manton \cite{KM}; its energy depends somewhat on the
$H$-to-$W$ mass ratio.  Moreover, a continuous set of static configurations
of $W$- and $H$-fields, leading from a vacuum at $N_{CS}=0$ to another at
$N_{CS}=1$ and passing through the sphaleron at $N_{CS}=1/2$, was
presented by Akiba, Kikuchi and Yanagida (AKY) \cite{AKY} (see also
ref.~\cite{BK}) who minimized the Yang--Mills and Higgs potential
energy for a
given value of $N_{CS}$.  This set of configurations describes a pass in the
potential energy surface over the Hilbert space of fields, the sphaleron
corresponding exactly to the saddle point of the pass. As most of the
workers in the field, we shall simplify the actual electroweak theory
by neglecting the Weinberg angle, i.e.~reducing it to the pure \SU case
(without the $U(1)$) and considering the simplest case of one Higgs
doublet. It should be mentioned that the sphaleron barrier has also been found
for non-zero Weinberg angle \cite{KKB}; however the difference
to the idealized case does not seem to be significant.

The purpose of this paper is to study fermion
field fluctuations around the whole sequence of
static YM--H configurations leading from $N_{CS}=0$ to 1.
To this end we investigate the behaviour of the Dirac sea of fermions as one
passes the sphaleron barrier. This is of great methodological interest, since,
as mentioned above,
the baryon and lepton numbers are changed during this passage
owing to the axial anomaly. We follow in detail this phenomenon.
The discrete level has previously
been studied in \cite{KB}. The complete Dirac spectrum, however, is calculated
for the first time in this work. We apply these results to determine the
fermionic contribution to the energy of the configurations on the path
from $N_{CS}=0$ to $N_{CS}=1$. This contribution is suppressed
by the factor $\alpha=g^2/4\pi\approx 0.04$ relative to the
classical one, but it also contains the large factor
$N_f$ which is $12$ for the real world. It is reasonable to consider the
range of parameters
$\alpha \ll \alpha N_f \ll 1$ where the fermionic energy is
a first order correction to the classical energy, but might have a
considerable numerical value.
In the case $\alpha N_f \sim 1$ one would have to solve
self-consistent equations for a sphaleron coupled to
fermions.

Having calculated the Dirac spectrum we evaluate the temperature
dependent corrections of the baryon number violation rate due to
the fermionic fluctuations.
For a realistic set of parameters $g=0.67$, $N_f=12$ ($2N_c+3=9$ massless
and $N_c=3$ massive doublets), $m_H=m_W$, 
and $m_t$ between $1.5\,m_W$ and $2.5\,m_W$ we find
numerically that the fermionic contribution
to the exponent of the Boltzmann factor increases the classical exponent
by up to $65\%$ in the temperature range $0.3\,m_W\leq T\leq 1.5\,m_W$,
which can considerably suppress the baryon
number changing sphaleron transition rate.
The suppression increases dramatically
with the top quark mass; at $m_t\approx 3\,m_W$ the fermionic
contribution to the exponent becomes as large as the classical one.

Finally our aim is to compute the baryon number and -density.
We confirm the law $B=N_{CS}$ and show
that for $N_{CS}\to 1$ the created baryon becomes a free, delocalized
particle.

The paper is organized as follows:
After recalling briefly the AKY \cite{AKY} configurations along the
barrier, we introduce the Dirac operator in the background field of those
configurations and diagonalize it in a spherical basis. We show how the
fermionic energy contributions can be extracted from the Dirac spectrum.
The energy of the Dirac sea is known to be divergent. Using
a proper time regularization scheme, we separate
the divergencies and combine them with the classical energy whose
parameters become renormalized.
We next consider the high temperature transition rate and derive the
fermionic part of its pre\-factor. The temperature range in which
the result can be applied will be discussed.
Finally we present our numerical results
and discuss them in the last section of the paper.

The simplified version of the standard electroweak
theory that we are going to study describes the \SU YM field
interacting with the $H$ doublet, and one left-handed fermion doublet
showing a Yukawa coupling to two right-handed singlets. In principle,
an \SU theory with an odd number of left-handed fermion doublets is
not properly defined because of the so-called global \SU anomaly \cite{W}. 
However, as we shall see
in practical terms we do not encounter any problems related to
that anomaly in the present work.

The Lagrangian is thus:
\begin{eqnarray}
{\cal L} &=& {\cal L}_{YMH} + {\cal L}_F, \nonumber \\
{\cal L}_{YMH} &=& -\frac{1}{4g^2}F_{\mu\nu}^a F^{a\,\mu\nu} 
+ (D_\mu \Phi)^\dagger
(D^\mu \Phi) - \frac{\lambda^2}{2}\left(\Phi^\dagger \Phi
-\frac{v^2}{2}\right)^2,
\la{LYMH} \\
{\cal L}_F &=& \bar{\psi}_L i\gamma^\mu D_\mu \psi_L +
      \bar{\chi}_R i\gamma^\mu \partial_\mu \chi_R
- \bar{\psi}_L M \chi_R - \bar{\chi}_R M^\dagger \psi_L
\la{LF}\end{eqnarray}
with $F_{\mu\nu}^a=\partial_\mu A_\nu^a - \partial_\nu A_\mu^a
      + \e^{abc}A_\mu^b A_\nu^c$
and the covariant derivative being defined as
$D_\mu=\partial_\mu -iA_\mu^a\tau^a/2$.
$M$ is a $2 \times 2$ matrix
composed of the Higgs field components $\Phi={\phi^+ \choose \phi^0}$,
and the Yukawa couplings $h_u, h_d$:
\beq
M=\matr{h_u \phi^{0\ast}}{h_d \phi^+}{- h_u \phi^{+\ast}}{h_d \phi^0}.
\la{MM}\eeq
$\psi_L$ means the \SU fermion doublet $u_L\choose  d_L$, and with $\chi_R$ we
denote the pair of the singlets $u_R$, $d_R$.
 
\section{AKY Configurations}
\setcounter{equation}{0}
\def\theequation{\arabic{section}.\arabic{equation}}

In order to minimize the energy of the static YM--H fields at a given
value of the Chern--Simons number,
\beq
N_{CS} = \frac{1}{16 \pi^2} \int d^3{\bf r}\;\e_{ijk}
\left(A_i^a\partial_jA_k^a+ \frac{1}{3}\e^{abc}A_i^aA_j^bA_k^c \right),
\la{CS}\eeq
one has to minimize the quantity
\beq
{\cal E} = E_{class} + 2 \xi N_{CS}\, .
\la{Lmult}\eeq
Here $\xi$ is a Lagrange multiplier and $E_{class}$ is the energy of the
static YM--H configuration which, in accordance with \eq{LYMH}, is
\beq
E_{class} = \int d^3{\bf r} \left\{ \frac{1}{2g^2}(B_i^a)^2 +
(D_i\Phi)^\dagger (D_i \Phi) +
\frac{\lambda^2}{2}\left(\Phi^\dagger \Phi
-\frac{v^2}{2}\right)^2 \right\},
\la{M}\eeq
where we have written $B_i^a=\frac{1}{2}\e_{ijk}F_{jk}^a$.
\par
At the classical level the $H$ field develops the v.e.v.
$\langle \phi^0 \rangle = v/\sqrt{2}$, $\langle \phi^+ \rangle = 0$,
and the $W$ boson field $A_\mu^a$ gets
a mass $m_W = gv/2$, while the $H$ field mass is $m_H = \lambda v$. We choose
the commonly used temporal gauge $A_0^a(x)\equiv 0$ and look
for the minimum of \eq{Lmult} in a spherically-symmetric "hedgehog" form
\cite{DHN,KM,AKY}:
\[
A_i^a({\bf r}) = \e_{aij}n_j\frac{1-A(r)}{r}
+ (\delta_{ai}-n_an_i)\frac{B(r)}{r}
+ n_an_i\frac{C(r)}{r}\,,
\]
\beq
\Phi({\bf r}) = \frac{v}{\sqrt{2}}\left[H(r)+i ({\bf n}{\bm\tau}) G(r) \right]
\doublet{0}{1}{}\;,
\la{HH}\eeq
where $r=|{\bf r}|$ and ${\bf n}={\bf r}/r$.
 
Gauge transformation which do not change the above
ansatz are given by
\beq
\Phi \rightarrow U\Phi, \;\;\; A_i \rightarrow
UA_iU^\dagger+iU\partial_iU^\dagger, \;\;\;
U = \exp[i({\bf n}{\bm\tau}) P(r)],
\la{GT}\eeq
with $A_i\equiv A_i^a\tau^a/2$.
Indeed, under this gauge transformation the functions 
$A$,$B$, $C$,$H$,$G$ transform as follows:
\beq
\begin{array} {l}
 A(r)\rightarrow A(r)\cos 2P(r) - B(r)\sin 2P(r), \\
 B(r)\rightarrow B(r)\cos 2P(r) + A(r)\sin 2P(r), \\
 C(r)\rightarrow C(r) + 2rP^\prime(r), \\
 H(r)\rightarrow H(r)\cos P(r) - G(r)\sin P(r), \\
 G(r)\rightarrow G(r)\cos P(r) + H(r)\sin P(r). \end{array}
\la{GTR}\eeq
This gauge freedom can be used to simplify the minimization of the
functional \ur{Lmult} for which we choose a gauge with $C(r)\equiv 0$. This
fixes the gauge transformation \ur{GT} up to a global rotation with a constant
$P$.

We next introduce dimensionless quantities,
\beq
x = r \frac{gv}{2} = r m_W, \;\;\;\; \kappa^2 = \frac{4\lambda^2}{g^2} =
\frac{m_H^2}{m_W^2}, \;\;\;\; \zeta = \frac{\alpha \xi}{2\pi m_W},
\la{dim}\eeq
and will measure the energy in units of $M_0=2\pi m_W/\alpha$. The
functional \ur{Lmult} becomes:
\begin{eqnarray}
&&\frac{{\cal E}}{M_0} = \frac{1}{2\pi}\int_0^\infty \! dx \left[A^{\prime 2}+
B^{\prime 2} + \frac{(A^2+B^2-1)^2}{2x^2}+2x^2(G^{\prime 2}+H^{\prime 2}) -
4BGH \right.
\nonumber \\
&&\qquad\quad\left. +2A(G^2-H^2)+(1+A^2+B^2)(G^2+H^2)+
\frac{\kappa^2}{2}x^2(G^2+H^2-1)^2 \right]  \nonumber\\
&&\qquad\quad +2\zeta N_{CS}\, .\la{Ed}
\end{eqnarray}
The Chern--Simons number $N_{CS}$ given by \eq{CS}
is only well defined for boson fields which are continuous and vanish
fast enough at infinity. In terms of the hedgehog ansatz \ur{HH} this 
means we have to use a gauge with $A(0)=A(\infty)=1$, $B(0)=B(\infty)=0$
and $C(0)=C(\infty)=0$, which is not fulfilled for fields 
with $C(r)\equiv 0$ which minimize the functional \ur{Ed}.
Hence, before inserting ansatz \ur{HH} into \eq{CS} we have
to gauge rotate the fields. A suitable transformation is obtained
from \eq{GTR} with $P(0)=-\varphi(0)/2$, $P(\infty)=-\varphi(\infty)/2$
where $\varphi(r)=\arg\left[A(r)+ iB(r)\right]$. By substitution
of the gauge rotated fields into \eq{CS} we obtain in terms of the
{\it unrotated} fields with vanishing $C$:
\beq
N_{CS} = \frac{1}{2\pi}\int_0^\infty \! dx\,(A^\prime B - B^\prime A) \;
+ \; \frac{1}{2\pi}\Bigl(\varphi(\infty)-\varphi(0)\Bigr).
\la{CSd}\eeq
Note that $N_{CS}$ does not depend on the detailed shape of $P(r)$. It is
therefore possible to perform the minimization of $\cal E$ 
with the unrotated fields $A$ and $B$ by inserting \ur{CSd} into \eq{Ed}.
Afterwards, the solution can easily be transformed into any other 
gauge by \ur{GTR}.

In order to find the minimum of ${\cal E}$ one has to solve the 
Euler--Lagrange eqs. for $A,B,G,H$ with boundary conditions compatible with the
finiteness of ${\cal E}$. The details are given in app.~A. One finds a
family of solutions labelled by $\zeta; \;\; -1<\zeta<1$. From \eq{CSd}
one then gets the relation between $N_{CS}$ and $\zeta$ which should be used to
express $E_{class}$ through $N_{CS}$. The values of $\zeta = -1, 0$ and $1$
correspond to $N_{CS}$ = 0, 1/2 and 1, respectively.

If one is interested in the case of finite fermion number density,
one has to add a term $2\mu N_{CS}$ to the energy functional \ur{M}, where
$\mu$ is the chemical potential for fermions \cite{R,DP}. Since the form of
the functional to be minimized is exactly the same as in the Lagrange
multiplier approach, used to find the minimal energy barrier
between the topologically distinct vacua, one proceeds in the same way as
above. The only difference is that in the case of $\mu \neq 0$ the energy
is $E_{class}(\mu)=E_{class}(0) + 2\mu N_{CS}$. We present the minimal energy
curves as a function of the Chern--Simons number for different values of
the chemical potential $\mu$ in Fig.~1. At $\mu = 0$ we
reproduce the curve of ref. \cite{AKY}. It can be seen that at $\mu =
\mu_{crit} = 2\pi m_W/\alpha$ the energy barrier disappears -- a fact which
was anticipated previously from analyzing vacuum stability
in the presence of finite baryon density \cite{R,DP}. At $\mu<\mu_{crit}$ 
the decay of a state
with finite baryon density is given by tunneling "bounce"
trajectories, found for small $\mu$ in ref.~\cite{DP}. It is a challenge 
for future work to
determine the decay rate of a high-density state for any $\mu$.
 
One should not be irritated by the cusps of the curve at integer $N_{CS}$:
the "coordinate" $N_{CS}$ is a quadratic functional of the gauge field, and in
a more natural coordinate the curve should be quadratic in the minima. To
illustrate this, we have considered a 2-dimensional $\sigma$ model with a
mass term, which mimics the 4-d electroweak theory. In this model the
minimal-energy configurations can be found analytically (D.Diakonov,
V.Petrov and M.Polyakov, unpublished), and the energy barrier is described
by the simple but exact formula $E = const \; |\sin \pi N_{CS} |$,
also exhibiting cusps at integer $N_{CS}$. However, if one sticks to just one
degree of freedom ($N_{CS}$), one should ask what is the kinetic energy along
this particular "coordinate". In the 2-d model this question can be
answered exactly, and the restricted kinetic energy turns out to be $\dot
N_{CS}^2 M(N_{CS})/2$ where the effective mass is $M(N_{CS}) = const\,/
| \sin \pi N_{CS} |$, being singular at integer $N_{CS}$. Therefore, if one 
would
replace $N_{CS}$ by a coordinate $X$ which would have a normal kinetic energy
$\dot X^2/2$, the potential energy would behave normally at the minima --
as $X^2$.
 
\section{Dirac Equation and Fermion Spectrum}
\setcounter{equation}{0}
\def\theequation{\arabic{section}.\arabic{equation}}

We are now starting to investigate the Dirac continuum and the fermion
discrete level in the background field of the static
minimal-energy configurations described in the previous section.

We define the Dirac Hamiltonian as the operator ${\cal H}$
appearing when one rewrites the Dirac equation following from \eq{LF}
in the form
\beq
\left(i \frac{\partial}{\partial t} - {\cal H}\right)\Psi = 0.
\la{DEQ}\eeq
Using the representation 
\beq
\gamma^0=\matr{1}{\ \,0}{0}{-1}\, ,\qquad
\gamma^i=\matr{0}{\sigma_i}{-\sigma_i}{0}\, ,\qquad
\gamma_5=\matr{0}{1}{1}{0}\, ,
\eeq
we can reduce the eight-spinors $\psi_L$, $\chi_R$ to four-spinors
$\tilde\psi_L$, $\tilde\chi_R$ by the definition
\beq
\psi_L\equiv\frac{1}{\sqrt{2}}{\tilde\psi_L\choose -\tilde\psi_L}\, ,\qquad
\chi_R\equiv\frac{1}{\sqrt{2}}{\tilde\chi_R\choose \tilde\chi_R}\, .
\eeq
In the basis of the four-spinors $\tilde\psi_L$, $\tilde\chi_R$ 
the Hamiltonian ${\cal H}$ is given by
\beq
{\cal H}=\matr{i\sigma_i D_i}{M}{M^\dagger}{-i\sigma_i\partial_i}
\la{fullham}\, ,\eeq
where $M$ is defined in \eq{MM}.
Here we have implied the $A_0\equiv 0$ gauge.

For the sake of numerical simplicity in finding the eigenvalues
of ${\cal H}$ we would like to exploit fully the spherical
symmetry of the AKY configurations. However, if the masses of the upper
components of the isospin doublets are different from the lower ones,
i.e. $h_u \neq h_d$, the spherical symmetry of the Dirac equation would be
spoiled. For that reason we shall only consider the case of equal
masses, $h_u = h_d = h$. This is certainly a very good approximation for
all leptons and quarks whose masses are much less than $m_W$ (which brings
the scale into the problem) but it is not so good for the three $(t,b)$
doublets. In Section~6 we will show how these massive doublets can be
treated in the framework of this approximation.

In the hedgehog ansatz \ur{HH} the matrices $M, M^\dagger$ have the
form:
\beq \begin{array}{c} M=m_F\left(H+i({\bf n}{\bm\tau})G\right) \\
M^\dagger=m_F\left(H-i({\bf n}{\bm\tau})G\right) \end{array}, \;\;\;
m_F=\frac{hv}{\sqrt{2}}\, .
\la{matrices}\eeq
To find the fermion levels one has to solve the eigenvalue problem:
\beq
{\cal H}\doublet{\tilde\psi_L}{\tilde\chi_R}{} = 
E\doublet{\tilde\psi_L}{\tilde\chi_R}{}\, ,
\la{eig}\eeq
where
\beq
{\cal H}=\matr{i\sigma_i D_i}{m_F\left(H+i({\bf n}{\bm\tau})G\right)}
{m_F\left(H-i({\bf n}{\bm\tau})G\right)}{-i\sigma_i\partial_i}.
\la{ham}\eeq
It is easy to check that the Dirac Hamiltonian ${\cal H}$ \ur{ham} commutes
with the so-called grand spin operator:
\beq
\op{K}=\op{L}+\op{S}+\op{T}\, ,
\la{GSO}\eeq
where $\op{L}$ is the angular momentum, $\op{S}$ is the spin, and $\op{T}$
is the isospin operator:
\beq
\hat{L}^a=-i\e^{abc}x_b\partial_c, \;\;\;
\hat{S}^a=\textstyle{\frac{1}{2}}\sigma^a, \;\;\;
\hat{T}^a=\textstyle{\frac{1}{2}}\tau^a,
\la{operators}\eeq
with $\op{S}^2=\op{T}^2=3/4$. Therefore, ${\cal H}$ can be diagonalized in
a basis of spherical harmonics with given grand spin $K$ and its third
component $K_3$. According to the coupling rules of angular momenta,
there are eight (in the case $K=0$ only four) basis vectors for fixed values of
$K$, $K_3$ and radial momentum $p$ (Both signs of the energy
are included). For the numerical diagonalization the basis
is made finite by discretization of the radial momentum; the allowed
values are denoted by $p_n$. For details the reader is referred to app.~B.
We call the matrix elements of ${\cal H}$ in this basis
${\cal H}_{KK_3}^{rs}(p_m,p_n)$, where $r,s=1\ldots 8$. Since
they do not depend on $K_3$ we suppress this index in
what follows; the degeneracy results in a factor $2K+1$. The eigenvalues
of ${\cal H}_K^{rs}(p_m,p_n)$ are denoted by $\e_{Kn}$ or $\e_\lambda$,
the ones of the free Dirac operator ${\cal H}^{(0)}$ are named
$\e_{Kn}^{(0)}$ or $\e_\lambda^{(0)}$,
given by $(\e_{Kn}^{(0)})^2=p_n^2+m_F^2$.
 
We have numerically calculated the eigenvalues $\e_{Kn}$ as a function of 
$N_{CS}$ for the transition from $N_{CS}=0$ to $N_{CS}=1$. The most interesting
behaviour is found in the $K=0$ sector, for which we plot some discretized
eigenvalues in Fig.~2. At $N_{CS}=0$ (vacuum), there are no discrete levels and
the spectrum is symmetric; each level is two-fold degenerate.
The levels of the lower continuum are occupied, the levels of the upper one
are empty. When $N_{CS}$ is increased, one level emerges from the lower
continuum, travels all the way through the fermion mass gap, and finally
reaches the upper continuum. This level stays occupied during the whole
transition. As can be seen from Fig.~2., the other levels are slightly
shifted upwards, each one replacing its predecessor. Eventually, at
$N_{CS}=1$ the spectrum looks exactly as at $N_{CS}=0$, except that
the first level of the upper continuum is now occupied, too.
This shows that the transition between topologically distinct vacua
is a baryon number violating process. Moreover, for intermediate 
(non-integer) values of $N_{CS}$ we have confirmed
numerically the law $B=N_{CS}$ following from the anomaly.
Both the bound state
level and the Dirac continuum contribute to the total baryon number
$B$ in this relation, and its proof requires an accurate gauge-invariant
regularization of the Dirac sea, see Section~6.
Previously, only the discrete level was investigated \cite{KB};
in this respect our results coincide with those of \cite{KB}.
 
Our next aim is to calculate the total energy contribution $E_{fer}$ of
the fermions. This energy is simply given by the sum of the energies of
all occupied states, relative to its value for the trivial vacuum
at $N_{CS}=0$. If we denote the energy of the discrete level which crosses zero
by $\e_{val}$, we obtain
\beq
E_{fer}=\sum_{\e_\lambda^{(0)}<0}\left(\e_\lambda-\e_\lambda^{(0)}\right)
=\sum_{\e_\lambda<0}\e_\lambda-\sum_{\e_\lambda^{(0)}<0}\e_\lambda^{(0)}
+\e_{val}\,\theta(\e_{val})\;,
\la{Efer} \eeq
with the step function $\theta$.
Since ${\rm Sp}\!\left({\cal H}-{\cal H}^{(0)}\right)=0$, we can also write
\beq
E_{fer}=-\frac{1}{2}\sum_\lambda\left(|\e_\lambda|-|\e_\lambda^{(0)}|\right)
+\e_{val}\,\theta(\e_{val})
\equiv E_{sea}+\e_{val}\,\theta(\e_{val})\;.
\la{Eseadef} \eeq
The sea energy $E_{sea}$ can also be written as
\beq
E_{sea}= -\frac{1}{2}\left({\rm Sp}\sqrt{{\cal H}^2}
  - {\rm Sp}\sqrt{{{\cal H}^{(0)}}^2}\right)
   =-\frac{1}{2}\sum_K (2K+1)\sum_n\left(|\e_{Kn}|-|\e_{Kn}^{(0)}|
    \right)\, .
\la{E}\eeq
The quantity $E_{fer}$ is in fact a fermion one-loop
correction to the classical energy of the
boson field, in particular it gives the quantum correction 
to the sphaleron
mass (another quantum correction arises from boson field fluctuations
around the sphaleron).
 
\section{Renormalization}
\setcounter{equation}{0}
\def\theequation{\arabic{section}.\arabic{equation}}

The aggregate energy of the Dirac sea \ur{E} is divergent, since
it is basically equivalent to the fermion loop in the external YM--H
field.  However, the electroweak theory is renormalizable which means that
all divergencies can be absorbed by local counterterms to the Lagrangian.
There are exactly four types of divergencies corresponding to four possible
counterterms:
\begin{enumerate}
\item a divergency which can be absorbed in the $B^2$ counterterm,
corresponding to the gauge coupling renormalization;
\item a divergency which can be absorbed in the $|D\Phi|^2$ counterterm,
corresponding to the Higgs "wave function" renormalization;
\item a divergency which can be absorbed in the $\lambda^2|\Phi|^4$
counterterm, corresponding to the Higgs quartic coupling renormalization;
\item a divergency which can be absorbed in the $m_H^2|\Phi|^2$ counterterm,
corresponding to the Higgs mass renormalization.
\end{enumerate}
The last divergency is quadratic in contrast to the logarithmic
divergencies in the first three cases.

Any renormalization assumes a certain "regularization scheme"; we shall use
a cut-off in a proper time representation for the Dirac sea
energy. This quantity can be written as
\beq
E_{sea}=-\frac{1}{2}\mbox{Sp}\;\sqrt{{\cal H}^2}=\frac{1}{4\sqrt{\pi}}
\int_\tau^\infty \frac{dt}{t^{3/2}}\;\mbox{Sp}\;\exp(-t{\cal H}^2),
\la{sqr}\eeq
where Sp is a functional trace, ${\cal H}$ is the Dirac
Hamiltonian \ur{ham}, and $\tau$ is the (inverse) ultra-violet cut-off, to
be taken eventually to zero. A non-zero $\tau$ suppresses Dirac levels with
high momenta, $|p| > 1/\sqrt{\tau}$. In \eq{sqr}
and below we do not write
the subtraction of the free Dirac sea explicitly, though we assume it.

\Eq{sqr} implies that the classical energy functional \ur{M} contains
bare couplings defined at the scale $1/\sqrt{\tau}$, such as the gauge
coupling $g(1/\tau)$, the Higgs expectation value $v(1/\tau)$, etc. 
However, when one finds e.g.~the classical mass of the sphaleron, 
one expresses it in
terms of the physical couplings and Higgs expectation value, defined at the
electroweak scale like $m_W$. Therefore, in order to use the standard
values for the physical constants, a proper time integral between $\tau$ and 
$1/m_W^2$ has to be absorbed by the classical energy. 
To be more precise, we write:
\beq
\int_\tau^\infty \frac{dt}{t^{3/2}}\;\mbox{Sp}\;\exp(-t{\cal H}^2) =
\int_\tau^{m_W^{-2}} + \int_{m_W^{-2}}^\infty .
\la{d1}\eeq
\indent
In the first integral one can perform a semi-classical expansion as it deals
with the spectral density of the operator ${\cal H}$ at high momenta. The
technique is explained e.g. in ref. \cite{DPY}. We get for the divergent
piece of the Dirac sea energy:
\[
\frac{1}{4\sqrt{\pi}}\int_\tau^{m_W^{-2}} \frac{dt}{t^{3/2}}\;\mbox{Sp}
\;\exp(-t{\cal H}^2)= \frac{1}{16\pi^2}\int_\tau^{m_W^{-2}}dt\int d^3{\bf r}
\left\{-\frac{2g^2m_F^2}{t^2m_W^2}(\Phi^\dagger\Phi-\frac{v^2}{2})\right.
\]
\[
\left.+\frac{1}{t}\left[\frac{1}{6}(B_i^a)^2
+ \frac{g^2m_F^2}{m_W^2} |D_i\Phi|^2 + \frac{g^4m_F^4}{2m_W^4}
\left((\Phi^\dagger\Phi)^2-\frac{v^4}{4}\right)\right]
+{\cal O}(t^0)\right\}
\]
\beq
\equiv
\frac{1}{4\sqrt{\pi}}\int_\tau^{m_W^{-2}}\frac{dt}{t^{3/2}}\left[\mbox{Sp}\;
\exp(-t{\cal H}^2)\right]_{div} + {\rm finite \; terms}.
\la{semicl}\eeq
\indent
We see that the
quadratically ($1/t^2$) and logarithmically ($1/t$) divergent terms are
exactly those entering the classical energy functional \ur{M}. Therefore,
they can and should be combined with the bare constants of the corresponding
terms in the classical energy -- to produce renormalized
physical constants at the scale of $m_W$. For example, the coefficients in
front of the magnetic field energy, $B^2/2$, combine into
\beq
\frac{1}{g^2(1/\tau)}-\frac{1}{48\pi^2}\ln (m_W^2\tau)=
\frac{1}{g^2(m_W^2)}\;,
\la{gcren}\eeq
and similarly for the other three terms. (\Eq{gcren} is not the whole
renormalization of the gauge constant since boson fluctuations also
contribute to it).
Because the divergent part of \ur{semicl} is combined with the
classical energy, the renormalized Dirac sea energy is:
\begin{eqnarray}
E_{sea}^{ren}&=&
\frac{1}{4\sqrt{\pi}}
\Biggl(\int_\tau^{m_W^{-2}}\frac{dt}{t^{3/2}}\;\mbox{Sp}\;\exp(-t{\cal H}^2) +
\int_{m_W^{-2}}^\infty\frac{dt}{t^{3/2}}\;\mbox{Sp}\;\exp(-t{\cal H}^2)
\nonumber \\
&&\qquad - \int_\tau^{m_W^{-2}}\frac{dt}{t^{3/2}}
\left[\mbox{Sp}\;\exp(-t{\cal H}^2)\right]_{div}\Biggr)\, . \la{renorm}
\end{eqnarray}
This expression is finite as $\tau\rightarrow 0$, and one can put
$\tau = 0$ in it. Introducing a dimensionless variable $tm_W^2
\rightarrow t$ and measuring all quantities including the sea energy
in units of $m_W$ we get:
\beq
E_{sea}^{ren} = \frac{1}{4\sqrt{\pi}}
\left(\int_0^\infty\frac{dt}{t^{3/2}}\;\mbox{Sp}\;\exp(-t{\cal H}^2) -
\int_0^1\frac{dt}{t^{3/2}}\left[\mbox{Sp}\;\exp(-t{\cal
H}^2)\right]_{div}\right)\,.
\la{ren}\eeq
\indent
Though \eq{ren} solves the problem of renormalization it is still
not too practical from the numerical point of view. The divergence of the
unrenormalized Dirac sea energy is due to the large contributions for
high values of $K$. Therefore it is convenient to perform the
renormalization subtraction for each $K$ sector separately, like
\beq
E^{ren}_{sea}=\sum_K(2K+1)(E_K-E_K^{div})\,.
\la{renK}\eeq
Then at each step
one deals with perfectly finite quantities, and the summation in
$K$ is convergent.

To this end we calculate the divergent part \ur{semicl} of the
functional trace (i.e the quantity to be subtracted) in a basis of
a complete set of eigenfunctions of the free Dirac Hamiltonian, given
in app.~B. From the definition of the trace we have:
\[
\frac{1}{4\sqrt{\pi}}\int_0^1\frac{dt}{t^{3/2}}\;\mbox{Sp}\;
\exp(-t{\cal H}^2)
=\sum_K(2K+1)\frac{1}{4\sqrt{\pi}}\int_0^1\frac{dt}{t^{3/2}}
\qquad\qquad \]
\beq \qquad\quad
\cdot\sum_{r=1}^8\sum_{n=1}^\infty
\int_0^R dx\,x^2
v_{K,i}^{(r)}(p_n,x)\left[\exp(-t{\cal H}_K^2)\right]_{ij}
v_{K,j}^{(r)}(p_n,x)\,.
\la{exactK}\eeq
where ${\cal H}_K$ is the Hamiltonian in the basis of the spherical
eigenfunctions for given grand spin $K$ and radial momentum $p_n$, in
fact an $8\times 8$ matrix, and $v_{K,i}^{(r)}(p_n,x)$ are the radial
parts of the eigenfunctions. For details see app.~B.

The semiclassical expansion in a given $K$ sector corresponds to expanding
the exponent in \eq{exactK} in powers of ${\cal H}_K-p_n$. The expansion is
rather tedious and the result is lengthy. We present the final
formula in app.~D. There we also show that
\eq{semicl} is reproduced from \eq{exactK} in the limit $R\to\infty$.
 
\section{Renormalization at Non-Zero Temp\-era\-tu\-res$^{\scriptscriptstyle
\rm 2}$}
\setcounter{equation}{0}
\def\theequation{\arabic{section}.\arabic{equation}}

\stepcounter{footnote}
\footnotetext{This
section was prepared in collaboration with V.Petrov and P.Pobylitsa.}
Keeping in mind the application to the baryon number changing sphaleron
transitions at non-zero temperatures (below the electroweak phase
transition) we explain here how the renormalization is performed for
the baryon number changing rate. This question has been addressed in
\cite{BochShap} and \cite{McLCar} for the limit of high temperatures;
it was demonstrated there how the theory is reduced to three dimensions.
In our work we will
not assume the high temperature limit but consider temperatures in the
range \ur{range} (see below). This leads to additional contributions
which are not present in the limit of high temperature.
 
In principle we treat both fermion and boson fluctuations on equal footing.
Our numerical calculations, however, are restricted to the corrections
due to fermion fluctuations. The bosonic part will be numerically dealt
with in a subsequent paper; so far it has been investigated in
e.g.~\cite{McLCar,McLetal,Baacke}. In an abelian $(1+1)$ dimensional
model it is possible to calculate the sphaleron transition rate
analytically. This has been done in \cite{Boch} for boson loop
corrections and in \cite{Gould} for fermions.

The traditional starting point is the semi-classical Langer--Affleck
formula \cite{L,A,KS} for the thermal transition rate $\Gamma$:
\beq
\Gamma = \frac{\beta\omega_{-}}{\pi}\;\mbox{Im}\,F,
\la{LA}\eeq
where $\omega_{-}$ is the negative boson eigenfrequency around the
sphaleron, $\beta$ is the inverse temperature and $F$ is the free energy of
the system.\footnote {A question may arise about the applicability of this
formula to the process under consideration since it implies that the
transition occurs mainly {\em above} the sphaleron barrier. It is not too
clear whether it is justifiable to treat such degrees of freedom as
belonging to thermal equilibrium, and whether presumably perturbative field
fluctuations above the barrier do actually lead to the baryon number
change. Keeping in mind these reservations we nevertheless start from
\eq{LA}.} One can imagine
introducing a small chemical potential for baryons so that the minima at
integer $N_{CS}$ become metastable, and $\mbox{Im}\,F$ gets a precise
meaning \cite{AM}.

The Langer--Affleck formula works in the temperature range given
approximately by \cite{A}
\beq
\frac{\omega_{-}}{2\pi}\le T\le E^{ren}_{class}(T)\, .
\la{range}\eeq
If the
temperature is lower than $\omega_{-}/2\pi$, the baryon number violation
rate is dominated by transitions via periodic solutions
of the classical Yang--Mills equations in Euclidean
space \cite{caloron}. In the case $T=0$ these solutions reduce to instantons.
Numerically the
value of $\omega_{-}/2\pi$ was found to be $\approx 0.3\,m_W$ for $m_H=m_W$
\cite{McLCar,AKYnew}. The
upper bound $E^{ren}_{class}(T)$ is the renormalized energy of the sphaleron
barrier. As we show below, it depends on $T$; it
vanishes at some critical temperature $T_c$ where the symmetry breaking of
the electroweak theory disappears. Hence, \eq{range} implies
$T\le T_c$. We will discuss numerical results in Section~6, but let us
mention here already that for $m_H=m_W$ and a reasonable value of the top 
quark mass, $T_c$ is of the order $m_W$ to $1.5\,m_W$. We see thus
that the domain of applicability of \eq{LA} is rather restricted, and the
reduction of the problem to three dimensions \cite{McLCar,McLetal,Baacke}
implying $T\gg m_W$ does not seem to be justified in reality.

In the semi-classical approximation one takes the static sphaleron
configuration as a saddle point for $\mbox{Im}\,F$, and has to compute
the boson and fermion determinants around it. Special care must be taken
towards the zero and negative modes of the boson small-oscillation
operator; this problem has been solved in ref.~\cite{McLCar,McLetal}, and we
denote the resulting factor by ${\cal N}_{0,-}$. The boson determinant below
involves only positive eigenmodes, and is denoted by a prime.
\Eq{LA} can be written as
\beq
\Gamma = \frac{\omega_{-}}{2\pi}\;{\cal N}_{0,-}
\Biggl(\frac{{\rm Det}_{bos}}{{\rm Det}_{bos}^{(0)}}\Biggr)
^{\prime\;-\frac{1}{2}}
\Biggl(\frac{{\rm Det}_{fer}}{{\rm Det}_{fer}^{(0)}}\Biggr)\exp
\left(-\beta E_{class}^{bare}\right)\, ,
\la{oneloop}\eeq
where "bare" means that the classical mass of the sphaleron is computed
with bare physical constants defined at some ultra-violet cut-off scale.
The superscript (0) refers to the free (no field) determinants.

Let us denote by $\omega_n^2$ the eigenvalues of the 3-dimensional
quadratic form for boson fluctuations about the sphaleron and by $\e_n$ the
eigenvalues of the Dirac Hamiltonian for fermions; the same quantities with
a superscript $0$ stand for the corresponding no-field eigenvalues.
Imposing boundary conditions periodic in time for bosons and antiperiodic
for fermions, one immediately establishes the eigenvalues of the
4-dimensional small-oscillation operators ($k\in\ZZ$):
\beq
{\lambda_{n,k}^{bos}}^2=\omega_n^2+\left(\frac{2k\pi}{\beta}\right)^2,
\;\;\; \lambda_{n,k}^{fer}=\e_n+i\frac{(2k+1)\pi}{\beta}\,.
\la{4eig}\eeq
The determinants of \eq{oneloop} are the products of these eigenvalues:
\[
\Biggl(\frac{{\rm Det}_{bos}}{{\rm Det}_{bos}^{(0)}}\Biggr)
^{\prime\;-\frac{1}{2}}=
\prod_n \prod_{k=-\infty}^\infty
\frac{\lambda_{n,k}^{bos,0}}{\lambda_{n,k}^{bos}} =
\prod_n\frac{\sinh(\omega_n^0\beta/2)}{\sinh(\omega_n\beta/2)}\, ,
\]
\beq
\Biggl(\frac{{\rm Det}_{fer}}{{\rm Det}_{fer}^{(0)}}\Biggr) =
\prod_n \prod_{k=-\infty}^\infty
\frac{\lambda_{n,k}^{fer}}{\lambda_{n,k}^{fer,0}} =
\prod_n\frac{\cosh(\e_n\beta/2)}{\cosh(\e_n^0\beta/2)}\, .
\la{dets}\eeq
At this point McLerran {\em et al.}~\cite{AM,McLCar,McLetal} argue that
in the case $T\gg m_W$, one should replace sinh by its argument and cosh
by unity,
since the 3-d eigenvalues $\omega_n, \;\e_n$ are of the order of $m_W$. We
will show later that this replacement is justified,
but as the condition $T\gg m_W$ is not compatible with the allowed
range \ur{range} we have to use the exact expressions in what follows.
The determinants \ur{dets} can be identically rewritten as:
\beq
\prod_n\frac{\sinh(\omega_n^0\beta/2)}{\sinh(\omega_n\beta/2)}=
\exp\left\{-\sum_n\frac{\beta}{2}(\omega_n-\omega_n^0)-
\sum_n\ln\left[\frac{1-e^{-\beta\omega_n}}{1-e^{-\beta\omega_n^0}}
\right]\right\},
\la{sh}\eeq
\beq
\prod_n\frac{\cosh(\e_n\beta/2)}{\cosh(\e_n^0\beta/2)}=
\exp\left\{\sum_n\frac{\beta}{2}(|\e_n|-|\e_n^0|)+
\sum_n\ln\left[\frac{1+e^{-\beta|\e_n|}}{1+e^{-\beta|\e_n^0|}}
\right]\right\}\, .
\la{ch}\eeq
The first terms in the exponents of \eqs{sh}{ch}
make up the zero-point oscillation corrections to the bare sphaleron mass
at zero temperatures:
\beq
E_{class}^{bare}+\sum_n\frac{1}{2}(\omega_n-\omega_n^0)-
\frac{1}{2}\sum_n (|\e_n|-|\e_n^0|) \equiv  E_1\, .
\la{msphal}\eeq
Both bosonic and fermionic sums are divergent but the divergencies
are absorbed in the renormalization of the constants entering
$E_{class}^{bare}$ so that $E_1$ is finite. In Section~4 this
renormalization was explicitly performed for the case of fermions.

The terms containing logarithms in \eqs{sh}{ch} depend on the temperature
but are ultra-violet finite, since the
large eigenvalues are cut by the Boltzmann factors. For a
closer inspection of these terms we use spectral densities,
in which the subtraction of the vacuum is included:
\beq
\sum_n\left[\ln (1-e^{-\beta\omega_n}) - \ln
(1-e^{-\beta\omega_n^0})\right] = \int_0^\infty dE
\,\rho^{bos}(E)\ln\left(1-e^{-\beta E}\right),
\la{rhobos}\eeq
\beq
\sum_n\left[\ln (1+e^{-\beta|\e_n|}) - \ln (1+e^{-\beta|\e_n^0|})
\right]
= \int_{-\infty}^\infty dE\,\rho^{fer}(E)\ln\left(1+e^{-\beta|E|}\right).
\la{rhoferm}\eeq
\indent
We can derive a semiclassical high--energy expansion for the density,
which is related to the expansion of \eq{semicl}. If the latter
one is written as
\beq
\mbox{Sp}\;\left(e^{-t{\cal H}^2}-e^{-t{{\cal H}^{(0)}}^2}\right)
=at^{-1/2}+bt^{1/2}+ct^{3/2}+...\; ,
\la{semiclpr}\eeq
one finds:
\begin{eqnarray}
\rho^{fer}(E)&=&|E|\;\mbox{Sp}\,\left(\delta({\cal H}^2-E^2)
-\delta({{\cal H}^{(0)}}^2-E^2)\right) \nonumber \\
&=&\frac{1}{\sqrt{\pi}}
\left(a-\frac{b}{2}\frac{1}{E^2}+\frac{3c}{4}\frac{1}{E^4}+...\right)\,.
\la{semiclE}\end{eqnarray}
We see that the first term corresponds to a constant spectral density
$\rho_\infty^{fer}$ at large $E$ (which, of course,
is in perfect correspondence
with the quadratic divergency of the Dirac energy). We get from \eq{semicl}
\beq
\rho^{fer}_\infty =\frac{a}{\sqrt{\pi}}
   = -\frac{g^2m_F^2}{2\pi^2m_W^2}\int d^3{\bf r}\left
(\Phi^\dagger\Phi-\frac{v^2}{2}\right)
\la{rhoinf}\eeq
and a similar expression (but with a different coefficient) for
$\rho_\infty^{bos}$. Putting these $\rho_\infty$ into \eqs{rhobos}{rhoferm}
we obtain for fermions:
\beq
\int_{-\infty}^\infty dE\,\rho^{fer}_\infty\ln\left(1+e^{-\beta|E|}\right)
=-\frac{g^2 m_F^2 T}{12m_W^2}
\int d^3{\bf r}\left(\Phi^\dagger\Phi-\frac{v^2}{2}\right)
={\cal O}\left(\frac{Tm_F^2}{m_W^3}\right)\, .
\la{Tren}\eeq
\indent
This expression is obtained for each fermion doublet so that we have
to sum over all doublets. Additionally, we must consider a
similar result for bosons, where $m_F^2$ has to be replaced by 
$(3/4)(m_H^2+3m_W^2)$ in the prefactor \cite{Kirz}.
At $T={\cal O}(m_W/g)$ these terms can be of the same order as the 
(renormalized)
classical zero-temperature energy of the sphaleron (divided by $T$),
therefore in that range of temperatures one has to find a new sphaleron
solution, with a new functional which includes the thermal part \ur{Tren}.
Fortunately, it has the same form as already encountered in the classical
energy functional (see \eq{M});
therefore, instead of solving anew the classical eqs.~of motion one can
use the non-thermal solution
for the YM--H fields but with the parameter $v$ replaced by the
temperature dependent expression
\beq
v(T)^2=v^2-\frac{3m_H^2+9m_W^2+4\sum_{\rm doubl.}m_F^2}{6m_H^2}\;T^2\;.
\la{vsubst}\eeq
This leads to a temperature dependent renormalization of the masses:
\beq
\frac{m_H(T)}{m_H}=\frac{m_W(T)}{m_W}=\frac{m_F(T)}{m_F}=\frac{v(T)}{v}\;.
\la{msubst}\eeq
With this thermal renormalization of the parameters
performed, one has to subtract the quantities $\rho_\infty^{bos,\,ferm}$
from $\rho^{bos,\,ferm}(E)$ in \eqs{rhobos}{rhoferm}. After the subtraction
high $E$ are suppressed as $1/E^2$ in the integrand so that in the limit
$T\gg m_W$ the contributions from \eqs{rhobos}{rhoferm} read:
\beq
\int_0^\infty dE \left(\rho^{bos}(E)-\rho_\infty^{bos}\right)\ln (\beta
E)\,,
\la{bos}\eeq
\beq
\int_{-\infty}^\infty dE \left(\rho^{fer}(E)-\rho_\infty^{fer}\right)
\ln 2\,.
\la{fer}\eeq
\indent
Apart from the subtleties with the renormalization mentioned above, this is
what one would naively get from \eq{dets} by replacing the fermionic cosh
by unity and the bosonic sinh by its argument, which corresponds
to the recipe of \cite{AM,McLCar,McLetal}. However, we are not working in
the high temperature limit but in the range \eq{range} so that we have
to use the full formula for the sphaleron transition rate:
\beq
\Gamma = \frac{\omega_{-}}{2\pi} {\cal N}_{0,-}
\exp \Bigl(-\beta E_{tot}(T)\Bigr),
\la{transrate}\eeq
where
\beq
E_{tot}(T)=E_{class}^{ren}(T)+E_{bos}^{temp}+E_{fer}^{temp}\, ,
\la{transsplit}\eeq
with
\begin{eqnarray}
&&\beta E_{bos}^{temp}=\beta \int_0^\infty dE
\left(\rho^{bos}(E)- \rho_\infty^{bos}-{\cal O}(1/E^2)\right) \frac{E}{2}
\nonumber \\
&&\qquad\qquad\quad +\int_0^\infty dE
\left(\rho^{bos}(E)-\rho_\infty^{bos}\right)\ln\left(1-e^{-\beta E}\right)
\nonumber\\
&&\beta E_{fer}^{temp}= -\beta \int_{-\infty}^\infty dE
\left(\rho^{fer}(E)- \rho_\infty^{fer}-{\cal O}(1/E^2)\right) \frac{|E|}{2}
\nonumber \\
&&\qquad\qquad\quad -\int_{-\infty}^\infty dE \left(\rho^{fer}(E)
   -\rho_\infty^{fer}\right)
\ln\left(1+e^{-\beta|E|}\right)\,.\la{transterms}
\end{eqnarray}
\indent
Here $E_{class}^{ren}(T)$ is the classical sphaleron mass with the
tempera\-ture-re\-norma\-lized values of the physical constants given by
\eqs{vsubst}{msubst}. According to
\eqs{transsplit}{transterms} the classical mass
gets corrections due to the temperature independent zero-point oscillations
(the first terms of $E_{bos}^{temp}$ and $E_{fer}^{temp}$)
and due to temperature dependent terms (the last terms of 
$E_{bos}^{temp}$ and $E_{fer}^{temp}$).
The ${\cal O}(1/E^2)$ terms to be subtracted from the spectral
densities correspond to the logarithmic divergencies written explicitly
for the fermions in \eq{semiclE}.

The boson loop corrections $\beta E_{bos}^{temp}$
are suppressed by the factor $\alpha$ so that they are usually discussed
in context with the pre-exponential factor. 
Its investigation is not performed in this work.
The fermionic contribution $\beta E_{fer}^{temp}$ is also suppressed by
$\alpha$, but it is enhanced by the number of fermionic species
$N_f=12$ so that its value can be considerable.
We shall see in next section that it suppresses
the transition rate significantly, especially for massive fermions.
 
\section{Numerical Results and Discussion}
\setcounter{equation}{0}
\def\theequation{\arabic{section}.\arabic{equation}}

In this section we present our numerical results.
Since the dependence on the Higgs mass $m_H$ turns out to be
weak we decided to fix $m_H=m_W$ for all calculations presented in this 
section. We will briefly comment on the influence
of $m_H$ at the end of the section. All results are
computed using the physical value $g=0.67$ for the coupling constant.

Let us start by discussing the behaviour of the fermion spectrum and the
contributions to the energy for the path from $N_{CS}=0$ to $N_{CS}=1$.
The level crossing phenomenon has already been demonstrated in Fig.~2.
Additionally, we have investigated
the discrete level for various fermion masses
$m_F$ and our results agree with those of Kunz and Brihaye \cite{KB}.

In the following we present numerical results for the fermionic energy
$E_{fer}$, especially
its sea part $E_{sea}^{ren}$, given by \eqs{Efer}{Eseadef} and \eq{ren}.
\Eq{ren} implies that the UV cut-off $\Lambda=1/\sqrt{\tau}$ has 
to be taken to infinity.
In practical terms, however, we always have to work with a finite $\Lambda$
since our numerical basis must be finite. Therefore we calculate
$E_{sea}^{ren}$ as
a function of $\Lambda$ and extrapolate to its value $E_{sea}^{ren}(\infty)$
at $\Lambda=\infty$.
From the semiclassical expansion of $E_{sea}$ for high $\Lambda$ we obtain
the following dependence:
\beq
E_{sea}^{ren}(\Lambda)=E_{sea}^{ren}(\infty)-\frac{C}{\Lambda^2}
+ {\cal O}\left(\frac{1}{\Lambda^4}\right)\, ,
\la{ecut}\eeq
with some constant $C$. (The terms $\sim\Lambda^2$ and $\sim\ln(\Lambda)$
are removed by renormalization.)
 
Apart from working with finite $\Lambda$ it is also necessary to
introduce two numerical parameters, the box size $R$ and the maximum
momentum $\Pmax$ (see app.~B), in order to obtain a finite basis for
the diagonalization of $\cal H$. For a fixed value of $\Lambda$ both have
to be taken large enough so that $E_{sea}^{ren}(\Lambda)$ does not
change any more when they are further increased. In Tab.~\ref{accut} we plot
results for fixed $m_F=2\,m_W$, $N_{CS}=0.5$ and $\Lambda=5\,m_W$
for different $R$ and $\Pmax$. 
\begin{table}[htbp]
\begin{center}
\begin{tabular}{|c||c|c|c|c|c|}
\hline
$\Pmax/\Lambda$ & 1.5  & 2.0 &  2.5 & 3.0 & 3.5 \\
\hline\hline
$R\,m_W=5$ & 7.98 & 7.29   &  7.20 &  7.20 & 7.20  \\
\hline
$R\,m_W=7$ & 8.99 &  8.34  & 8.25  &  8.25 &  8.25 \\
\hline
$R\,m_W=9$ & 9.25 &  8.57  & 8.49  &  8.48 &  8.48 \\
\hline
$R\,m_W=10$ & 9.25 & 8.61   & 8.52  & 8.52  &  8.52 \\
\hline
$R\,m_W=11$ & 9.28 &  8.62  &  8.54 & 8.54  &  8.54 \\
\hline
$R\,m_W=12$ & 9.31 &  8.64  &  8.55 &  8.55 & 8.55  \\
\hline
$R\,m_W=13$ & 9.31 & 8.64   &  8.55 & 8.55  & 8.55  \\
\hline
$R\,m_W=14$ & 9.28 &  8.64  &  8.55 &  8.55 & 8.55  \\
\hline
\end{tabular}
\parbox{5in}{\caption{\it
$E_{sea}^{ren}(\Lambda=5\,m_W)$ for different values of the
numerical parameters $R$ and $\Pmax$ with $m_F=2\,m_W$, $N_{CS}=0.5$
and $\Lambda=5\,m_W$ fixed. Stability is reached at $R\approx 12\,m_W^{-1}$
and $\Pmax\approx 3\,\Lambda$.
}
\la{accut}}
\end{center}
\end{table}
It can be seen that stability is reached
for $R\approx 12\,m_W^{-1}$ and $\Pmax\approx 3\,\Lambda$ so that we obtain as
our result:
\beq
E_{sea}^{ren}(\Lambda=5\,m_W,m_F=2\,m_W,N_{CS}=0.5) = 8.55\,m_W\, ,
\eeq
with a numerical error of less than $0.2\,\%$.
Using the same method we obtain results for other values of $\Lambda$
and for different $m_F$ and $N_{CS}$. It turns out that in general
it is sufficient to choose $R=12\,m_W^{-1}$ and $\Pmax=3.5\,\Lambda$. In
Tab.~\ref{aclam} we present results for various values of $\Lambda$ with
$R=12\,m_W^{-1}$ fixed. Additionally we have written values for
$E_{sea}^{ren}(\Lambda)$ obtained from \eq{ecut} with
$E_{sea}^{ren}(\infty)= 9.09\,m_W$ and $C=13.54\,m_W^3$ into the last
column. 
\begin{table}[htbp]
\begin{center}
\begin{tabular}{|c||c|c|c|c|c||c|}
\hline
$\Pmax/\Lambda$ & 1.5  & 2.0 &  2.5 & 3.0 & 3.5 & from \eq{ecut}\\
\hline\hline
$\Lambda/m_W = 3$ & 8.28 & 7.71   & 7.64  & 7.63  & 7.63 &  7.59 \\
\hline
$\Lambda/m_W = 4$ & 8.86 &  8.33  &  8.25 & 8.25  & 8.25 &  8.24 \\
\hline
$\Lambda/m_W = 5$ & 9.27 &  8.63  & 8.55  & 8.55  & 8.55 &  8.55 \\
\hline
$\Lambda/m_W = 6$ & 9.54 &  8.80  & 8.71  & 8.71  & 8.71 &  8.71 \\
\hline
$\Lambda/m_W = 8$ & 9.89 & 8.98   & 8.88  & 8.88  & 8.88 &  8.88 \\
\hline
\end{tabular}
\parbox{5in}{\caption{\it
$E_{sea}^{ren}(\Lambda)$ for various values of $\Lambda$ and
$\Pmax$. $R=12\,m_W^{-1}$, $m_F=2\,m_W$ and $N_{CS}=0.5$ are fixed. The values
in the last column are obtained by eq.~(6.1) with
$E_{sea}^{ren}(\infty)=9.09\,m_W$ and $C=13.54\,m_W^3$.
}
\la{aclam}}
\end{center}
\end{table}
A comparison with the numerical results shows that
the power law \eq{ecut} is fulfilled very accurately. Therefore
the numerical error of the extrapolated value
$E_{sea}^{ren}(\infty)$ is
of the same order as the error of  $E_{sea}^{ren}(\Lambda)$
for finite $\Lambda$. Thus our extrapolated result 
for $E_{sea}^{ren}$ finally reads:
\beq
E_{sea}^{ren}(m_F=2\,m_W,N_{CS}=0.5) = 9.09\,m_W\, ,
\eeq
again with a numerical error of $0.2\,\%$.
Our results for different values of $m_F$ and $N_{CS}$ were obtained
the same way. The numerical errors always have roughly the same size.
The relative error can
increase up to $2\%$ for small $m_F$ where $E_{sea}^{ren}$ is considerably
lower than for $m_F=2\,m_W$.
This accuracy is fully sufficient for our purpose; one should keep
in mind that
the error arising from taking equal masses
for the top and bottom quark is surely bigger than the numerical error.

In Fig.~3 we have plotted $E_{fer}^{ren}\equiv E_{sea}^{ren}
+\e_{val}\,\theta(\e_{val})$
as a function of $N_{CS}$ for $m_F/m_W=0$, $1$ and $2$. For
$m_F=0$ there is no contribution from the valence level, the energy is
symmetric
with respect to $N_{CS}=0.5\,$. Its total value is very small ($E_{sea}^{ren}=
0.18\, m_W$ for $N_{CS}=0.5$) compared to the classical energy of the
boson fields
($\approx 100\, m_W$ for $N_{CS}=0.5$). This is also the case for $m_F=m_W$,
where the
energy is the sum of the symmetric sea and antisymmetric valence part.
The fermionic energy $E_{fer}^{ren}$ is only significant in comparison to
the classical energy when $m_F$ is much larger than $m_W$.
For $m_F=2\,m_W$ the
energy of the fermions is already about $9\%$ of the classical energy, and
it increases dramatically with $m_F$. In fact, one can show that in the
leading
order the sea energy grows with $m_F$ as $m_F^4\ln(m_F)$. This law is already
a good approximation for $m_F\geq 3\, m_W$. In Tab.~\ref{grow} we have plotted
results of $E_{sea}^{ren}$ for different $m_F$ at $N_{CS}=0.5\,$.
\begin{table}[htbp]
\begin{center}
\begin{tabular}{|c|c|c|c|c|c|c|c|}
\hline
$m_F/m_W$ &0 & 1& 2  & 3  & 4   & 5 & 6  \\
\hline
$E_{sea}^{ren}/m_W$ &0.18& 0.51 & 9.05 & 72.63 & 303.5 & 862 & 2001 \\
\hline
\end{tabular}
\parbox{5in}{\caption{\it
$E_{sea}^{ren}$ as a function of $m_F$ for fixed $N_{CS}=0.5\,$.
The energy increases according to $E_{sea}^{ren}\sim m_F^4\ln(m_F)$.
It should be compared to the classical energy which is about $100\,m_W$
for $N_{CS}=0.5\,$.
}
\la{grow}}
\end{center}
\end{table}
 
Before we turn to the discussion of the
temperature dependent contribution $E_{fer}^{temp}$ to the sphaleron 
transion rate and the baryon number and -density, let us briefly
comment on the numerical accuracy of these UV-finite quantities.
The only parameters which turn up in the numerical evaluation are
$R$ and $\Pmax$. Again, both have to be chosen large enough to
ensure stability of the results. This is done in analogy to
the method for the calculation of $E_{sea}^{ren}(\Lambda)$ for fixed
$\Lambda$. The numerical errors are again in the range of $1\%$ to
$2\%$.

We are now going to present the temperature dependent contributions to the
sphaleron transition rate, given by \eqsss{transrate}{transterms}. In this 
paper
we only considered the classical term $\beta E_{class}^{ren}$ and the
fermionic part $\beta E_{fer}^{temp}$ but not the bosonic part 
$\beta E_{bos}^{temp}$.
The temperature enters the transition rate in three ways: First, in
form of the prefactor $\beta=1/T$ which causes the transition rate to
become unsuppressed when $E_{tot}(T)\sim T$. Second, the masses get
renormalized via \eq{msubst}
so that the energy $E_{tot}(T)$ is roughly proportional to $v(T)/v(0)$
given by \eq{vsubst}. Third, the last term of $E_{fer}^{temp}$ is
explicitly dependent on $\beta$. The first two effects are
numerically trivial but very large, while the last one 
is difficult to compute and rather small.

We start by considering the first two aspects. The dominant
contribution in \eq{transsplit} is
\beq
\beta E^{ren}_{class}(T)=\frac{1}{T} E^{ren}_{class}(0)
\left(1-\frac{3m_H^2+9m_W^2+4\sum_{\rm doubl.}m_F^2}{6v^2m_H^2}\,T^2\right).
\la{etotT}\eeq
For $T\to 0$ this terms goes to infinity, but it decreases rapidly with
increasing $T$. In the vicinity of the critical temperature
\beq
T_c^2=\frac{6v^2m_H^2}{3m_H^2+9m_W^2+4\sum_{\rm doubl.}m_F^2}\,,
\la{Tcrit}\eeq
where the symmetry breaking of electroweak theory and the sphaleron
barrier disappear the function $\beta E^{ren}_{class}(T)$
goes to zero and the baryon number violation rate
becomes unsuppressed.

The value of $T_c$ and the quantitative behaviour of
$\beta E^{ren}_{class}(T)$ depend largely on the choice of the
parameters. We wish to focus our interest to a situation which resembles
the physical one as close as possible.
As it was mentioned above to preserve spherical symmetry of the Dirac
equation one has to consider the case of equal masses for up and down
fermions. This approximation is not so good for the three
$(t,b)$ doublets. In the case $m_{t,b} < m_W$ the fermionic corrections
can be estimated using the average mass of the doublet, and treating the
difference as a perturbation. The corrections are anyhow small in this
case. In the more realistic case $m_b\ll m_W \ll m_t$ the correction
from the top quark is {\it half} that of the doublet with both masses
equal to $m_t$ (This recipe can easily be derived from eqs.~\ur{renorm}
and \ur{semiclE} in the limit $m_t\gg m_W$). Therefore, in our numerical
estimates we take $9+3/2$ massless fermion doublets and $3/2$ massive
doublets with mass $m_t$ which we vary in the range $m_t/m_W=1.5,\,2.0,\,
2.5$. In particular the sums in \eqs{etotT}{Tcrit} are replaced by
$3/2\,m_t^2$.

We have plotted $\beta E^{ren}_{class}(T)$ as a function of $T$
in Fig.~4 (dashed lines) for the masses $m_t/m_W=1.5$, $2.0$ and $2.5$.
We realize that $T_c$ is rather low (between
$m_W$ and $1.5\,m_W$). One can obtain a higher $T_c$ by taking a larger
Higgs mass, but this will not change the picture qualitatively. The current
experimental bounds do not suggest that $m_H$ should be considerably higher
than $m_W$. We have already stated in Section~5 that the range of applicability
of the Langer--Affleck formula is roughly given by $0.3\,m_W\approx
\omega_{-}/2\pi\le T\le T_c\approx 1.5\,m_W$. This is the range we use
in Fig.~4, but one has to be careful if $T$ is very close to $T_c$. In this
case the Langer--Affleck formula is clearly not valid because the barrier gets
so low that the transition occurs far above the sphaleron.

We now want to investigate how far the fermionic fluctuations
$\beta E_{fer}^{temp}$ of \eq{transterms} influence the transition rate.
The numerical results for this contribution are included in Fig.~4
(solid lines), again for the three choices $m_t/m_W=1.5$, $2.0$ and $2.5$.
We find that the results for these cases are quite different. 
For $m_t=2.5\,m_W$ and $T\approx 0.5\,m_W$ the fermionic fluctuations
increase the exponent of the classical
Boltzmann factor by about $50\%$. For $T\approx 0.9\,m_W$
this increase is even about $65\%$.
On the other hand, for $m_t=1.5\,m_W$ the fluctuations are only between
$5$ and $9\%$ of the classical exponent.
We conclude that the fermion loops suppress the baryon
number violation rate in any case, but the strength of the suppression
depends on the value for the mass of the top quark. We expect that the above
numbers slightly change if the difference between $m_t$ and $m_b$ is
treated perturbatively in higher orders.

The suppression of the transition rate is important for
the understanding of the matter excess in the universe. However, such
cosmological consequences are still a matter of discussion
(see e.g.~\cite{Shap1,Shap2}), and we will not go into details here.

Let us turn to the discussion of the baryon number and -density. First we have
to find reasonable definitions for these quantities. In principle, the baryon
number is the number of all negative energy eigenstates of the Dirac
Hamiltonian, relative to the vacuum. The valence state always contributes to
the baryon number, so it has to be added explicitly after it has crossed zero.
However, the increase of the baryon number according to the law $B=
N_{CS}$ is not only caused by the level crossing of the valence state,
but it is also due to a process that happens at both ends of the
spectrum. To see this let us return to Fig.~2 which demonstrates
that, going from $N_{CS}=0$ to $N_{CS}=1$, one state emerges from
''outside the spectrum'' and joins the negative continuum, and
another one disappears ''from the positive continuum to infinity''.
In order to take this effect into account correctly we have to introduce
a regularization and compute the result with a finite cut-off which
eventually must be sent to infinity. Thus we define:
\begin{eqnarray}
B&=&\lim_{\Lambda\to\infty}\left\{N\bigl(\e_\lambda<0\bigr)
-N\bigl(\e_\lambda^{(0)}<0\bigr)\right\}_{reg}+\theta(\e_{val})
 \nonumber\\
&=&\lim_{\Lambda\to\infty}\left\{-\frac{1}{2}\sum_\lambda
  {\rm sign}\left(\e_\lambda\right)\right\}_{reg}
  +\theta(\e_{val}) \nonumber \\
&=&\lim_{\Lambda\to\infty}\left\{-\sum_\lambda\frac{\e_\lambda}
  {2\sqrt{\pi}}\int_{\frac{1}{\Lambda^2}}^\infty dt\,t^{-1/2}
  \, e^{-\e_\lambda^2t}\right\}+\theta(\e_{val})\, . \la{barnum}
\end{eqnarray}
It is possible to show analytically that $\delta B=\delta N_{CS}$
with $N_{CS}$ and $B$ being defined by \eq{CS} and \eq{barnum}, respectively.
Here ''$\delta$'' means the variation with respect to the boson fields. 
One has to differentiate the integral in \eq{barnum} with respect
to $\e_\lambda$, perform the $t$ integral and expand in powers of
$1/\Lambda$. The only term which does not vanish in the limit $\Lambda
\to\infty$ is identical to $\delta N_{CS}$. For the baryon density we
obtain:
\[
\rho(r)=\lim_{\Lambda\to\infty}\int d\Omega\sum_\lambda
\varphi_\lambda^\dagger({\bf r})\varphi_\lambda({\bf r})
\left[-\frac{\e_\lambda}{2\sqrt{\pi}}\int_{\frac{1}{\Lambda^2}}^\infty
dt\,t^{-1/2}\, e^{-\e_\lambda^2t}\right]
\]\beq
+\theta(\e_{val})\int d\Omega\,\varphi_{val}^\dagger({\bf r})
\varphi_{val}({\bf r})\, .
\la{bardens}\eeq
Here $\varphi_\lambda({\bf r})$ are the 
eigenfunctions of the Dirac Hamiltonian:
${\cal H}\varphi_\lambda=\e_\lambda\varphi_\lambda$. 
Of course, by definition we
have $B=\int_0^\infty dr\,r^2\rho(r)$.

In order to evaluate \eqs{barnum}{bardens} numerically we have to insert
the eigenvalues $\e_\lambda$ obtained from the diagonalization in different
sectors of grand spin $K$. It turns out that $B$ and $\rho$ are dominated by
the $K=0$ sector. In fact, its contribution is usually more than $90\%$ of the
total value. This means that the created baryon is basically in a $K=0$ state.

In Tab.~\ref{bartab} we present the baryon number $B$ as a function of
$N_{CS}$ for $m_F=m_W$.
\begin{table}[htbp]
\begin{center}
\begin{tabular}{|c|c|c|c|c|c|c|c|c|c|c|c|}
\hline
$N_{CS}$ & 0.00  & 0.15& 0.26& 0.35 & 0.43 & 0.50 & 0.57 & 0.65 &
   0.74 & 0.85 & 1.00 \\
\hline
$B$ & 0.00 & 0.18 & 0.27 & 0.35 & 0.43 & 0.50 & 0.57 & 0.65 &
    0.73 & 0.82 & 1.00 \\
\hline
\end{tabular}
\parbox{5in}{\caption{\it $B$ as a function of $N_{CS}$ for $m_F=m_W$.
The data confirm the law $B=N_{CS}$ coming from the anomaly.}
\la{bartab}}
\end{center}
\end{table}
We observe that the law $B=N_{CS}$ is excellently
reproduced. The same behaviour is obtained for different masses $m_F$.
The baryon density as a function of the radial distance $r$ is plotted
in Fig.~5 for $N_{CS}=0.26$, $0.5$, $0.74$ and close to $1$
($m_F=m_W$). Between $N_{CS}=0$ and $N_{CS}=0.5$ the density increases
with $N_{CS}$ for each fixed $r$. For $N_{CS}>0.5$ the density at the
origin $\rho(0)$ decreases
again, while the integral $B$ is still increasing (as seen from
Tab.~\ref{bartab}).
Thus we are creating a baryon which becomes more and
more delocalized. Approaching $N_{CS}=1$ the density becomes independent
of $r$ and infinitesimally small everywhere. This corresponds to
a free, unbound particle as expected from a vacuum configuration
of the classical fields.

Finally we have checked the influence of the Higgs mass $m_H$ on the
preceding results. Qualitatively, the picture was found to be the
same for any $m_H$ (at very large $m_H$, however, the sphaleron solution
becomes modified, see \cite{BK}).
In fact, we observe that both $E_{fer}^{ren}$
and $E_{fer}^{temp}$ are slightly decreasing functions of $m_H$.
So for $m_H>m_W$
they always have the same order of magnitude as their classical
counterparts. Only for the unphysical value $m_H\to 0$ the fermionic
terms show a significant increase. In this case the energy  
$E_{fer}^{ren}$ reaches approximately $10\%$ of the classical 
energy ($m_F=m_W$),
which is about 20 times higher than for $m_H=m_W$.
This relatively
large value is due to the slow decrease of the boson fields for
$r\to\infty$ which causes a strong perturbation of the Dirac sea.

{\bf Acknowledgement:}
We are grateful to V.Petrov and P.Pobylitsa for numerous discussions.
D.D.~and M.P.~would like to thank the
Institute for Theoretical Physics II
of the Ruhr Universit\"at Bochum for hospitality.
Their work was sponsored in part by the Russian Foundation for
Fundamental Research under grant $\#$93-02-3858 and by the
Deutsche Forschungsgemeinschaft.
\appendix
\section{A}
\setcounter{equation}{0}
\def\theequation{\Alph{section}.\arabic{equation}}

For the sake of completeness we cite the eqs.~for getting the
minimal-energy configurations of Akiba {\it et al.} \cite{AKY}.
They are found from the Euler--Lagrange eqs.~which follow from varying
\eq{Ed}:
\begin{eqnarray}
A^{\prime\prime}&=&\frac{1}{x^2}A(A^2+B^2-1)
   +A(H^2+G^2)+G^2-H^2-2\zeta B^\prime,
\nonumber \\
B^{\prime\prime}&=&\frac{1}{x^2}B(A^2+B^2-1)+B(H^2+G^2)-2HG+2\zeta A^\prime,
\nonumber \\
(xH)^{\prime\prime}&=&\frac{1}{2x}H[(A-1)^2+B^2]-\frac{1}{x}BG+
\frac{\kappa^2}{2}xH(H^2+G^2-1),
\nonumber \\
(xG)^{\prime\prime}&=&\frac{1}{2x}G[(A-1)^2+B^2]+\frac{2}{x}AG-\frac{1}{x}BH+
\frac{\kappa^2}{2}xG(H^2+G^2-1). \nonumber \\
\la{eqs}\end{eqnarray}
A consequence of these eqs.~is a "conservation law" which can be used to
check the accuracy of the numerical integration:
\beq
(AB^\prime - A^\prime B) + x^2 (HG^\prime - H^\prime G)+\zeta(1-A^2-B^2)=0.
\la{CL}\eeq
\indent
The boundary conditions for eqs.~\ur{eqs} follow from the requirement that
there are no singularities at the origin and that the energy functional
does not diverge at infinity. The first requirement implies
the following behaviour of the functions near the origin:
\begin{eqnarray}
A&=&\cos\alpha\,(1+rx^2)
    -\fracsm{1}{3}(st+2r\zeta)(\sin\alpha)\,  x^3 + ...\; ,
\nonumber \\
B&=&\sin\alpha\,(1+rx^2)+\fracsm{1}{3}(st+2r\zeta)(\cos\alpha)\,x^3 + ...\; ,
\nonumber \\
H&=&s\,\cos\fracsm{\alpha}{2}+t(\sin\fracsm{\alpha}{2})
\,x+\fracsm{1}{12}\kappa^2 s(s^2-1)
  (\cos\fracsm{\alpha}{2})\,  x^2 + ... \; ,
\nonumber \\
G&=&s\,\sin\fracsm{\alpha}{2}-t(\cos\fracsm{\alpha}{2})
\,x+\fracsm{1}{12}\kappa^2 s(s^2-1)
  (\sin\fracsm{\alpha}{2})\,  x^2 + ... \; ,
\la{or}\end{eqnarray}
where $\alpha, r,s,t$ are arbitrary constants.

Requiring the convergence of the energy integral at infinity and solving
the eqs.~\ur{eqs} at large $x$ one gets the following behaviour of the
functions at $x\rightarrow\infty$:
\begin{eqnarray}
A&=&\cos\gamma + e \sin(\zeta x + \beta) \; E(x)\,,
\nonumber \\
B&=&\sin\gamma - e \cos(\zeta x + \beta) \; E(x)\,,
\nonumber \\
H&=&\cos\fracsm{\gamma}{2}\left(1+\frac{h}{x}e^{-\kappa x}\right)
-e\sin\fracsm{\gamma}{2}\cos(\zeta x +\beta -\gamma +\delta)\;
\frac{E(x)}{x^2}\,,
\nonumber \\
G&=&\sin\fracsm{\gamma}{2}\left(1+\frac{h}{x}e^{-\kappa x}\right)
+e\cos\fracsm{\gamma}{2}\cos(\zeta x +\beta -\gamma +\delta)\;
\frac{E(x)}{x^2}\,,
\la{inf}\end{eqnarray}
where we have denoted
\beq
E(x) \equiv \exp \left(-\sqrt{1-\zeta^2}x\right),\;\;\;
\zeta = \sin\fracsm{\delta}{2}\,,\;\;\; \sqrt{1-\zeta^2} =
\cos\fracsm{\delta}{2}\, ,
\la{not}\eeq
and $\gamma$, $\beta$, $e$ and $h$ are arbitrary constants.

Note that the asymptotics \ur{inf} is valid only if $m_H \sim m_W$; if
the Higgs is much heavier than the $W$ boson, eqs.~\ur{inf} have
to be modified.

We have thus four arbitrary constants defining the behaviour of
the functions at infinity ($\beta,\gamma,e,h$) and four in the origin
($\alpha,r,s,t$). Matching solutions of eqs.~\ur{eqs} starting from the
origin with those starting from infinity, one finds these constants for
a given value of $\zeta$ (and $\kappa$) -- up to a global gauge rotation,
however. The point is, only the combination $\gamma-\alpha$ is invariant
under the gauge rotation \ur{GTR} with a constant $P$. One can use this
freedom to fix $\alpha =0$, for example.

It should be noticed that eqs.~\ur{eqs} have a solution only for
$|\zeta|< 1$. The end points of this interval correspond to $N_{CS} = 0,\;1$;
it is seen from \eqs{inf}{not} that at $|\zeta|\rightarrow 1$ the functions
rather oscillate than decrease.
 
\section{B}
\setcounter{equation}{0}
\def\theequation{\Alph{section}.\arabic{equation}}

In this appendix we summarize how
the Dirac Hamiltonian ${\cal H}$ of \eq{ham}
is diagonalized. Since it commutes with the grand spin operator
$\op{K}$, we diagonalize it in a basis of
eigenfunctions of $\op{K}^2$ and
$\opr{K}_3$.
To this end we introduce normalized eigenstates
of the commuting operators $\op{L}^2, \op{J}^2 =
(\op{L}+\op{S})^2, \op{K}^2 = (\op{J}+\op{T})^2$ and its third
component $\opr{K}_3$. Naturally, they are also
eigenstates of the operators $\op{S}^2=\op{T}^2=3/4$.
We shall use the short-hand notation $\ket{K,J,L;K_3}$
to denote these states.

Since only the operators $\op{K}^2$, $\opr{K}_3$ commute with
the Hamiltonian ${\cal H}$
but not $\op{J}^2$ or $\op{L}^2$, states labelled by different $J, L$
at given $K$, $K_3$ will mix in the eigenvalue \eq{eig}.
Generally speaking,
for given values of $K$, $K_3$ the following four states mix:
\[
\ket{K,\:K+\frac{1}{2},\:K+1;K_3},\;\;\; \ket{K,\:K+\frac{1}{2},\:K;K_3},
\]
\beq
\ket{K,\:K-\frac{1}{2},\:K;K_3},\;\;\; \ket{K,\:K-\frac{1}{2},\:K-1;K_3}.
\la{states}\eeq
We therefore have to decompose the spinor-isospinor functions $\tilde\psi_L,
\tilde\chi_R$ in the above four states for every value of $K$, $K_3$:
\begin{eqnarray}
\tilde\psi_L({\bf r})&=&\sum_{K=0}^\infty\sum_{K_3=-K}^K
\sum_{J=K-\frac{1}{2}}^{K+\frac{1}{2}}
\sum_{L=J-\frac{1}{2}}^{J+\frac{1}{2}} i^L\:f_{JL}^{KK_3}(r)
\:\langle\Omega\ket{K,J,L;K_3}\, ,\nonumber \\
\tilde\chi_R({\bf r})&=&\sum_{K=0}^\infty \sum_{K_3=-K}^K
\sum_{J=K-\frac{1}{2}}^{K+\frac{1}{2}}
\sum_{L=J-\frac{1}{2}}^{J+\frac{1}{2}} i^L\:g_{JL}^{KK_3}(r)
\:\langle\Omega\ket{K,J,L;K_3}\, . \la{dec}
\end{eqnarray}
\indent
The eigenvalue \eq{eig} is
thus mixing four functions $f$ (from $\tilde\psi_L$)
and four functions $g$ (from $\tilde\chi_R$) for any value of $K$, $K_3$.
Since the Hamiltonian is block-diagonal with respect to $K_3$,
\eq{eig} results in a system of eight ordinary differential eqs.~for eight
independent functions for each $K$. In the case of $K=0$ where the last
two states in \eq{states} are absent, only two states for $\tilde\psi_L$
and two for $\tilde\chi_R$ mix, and the eigenvalue equation involves only
four functions. This particular case was considered recently in
ref.~\cite{KB}.

The set of differential equations will be solved by putting the system into
a large but finite spherical box. To insure that at the boundaries of the box
the boson fields take their vacuum values, we have to switch to a
gauge in which $A(\infty)=1$, $B(\infty)=0$, $C(\infty)=0$, $H(\infty)=1$
and $G(\infty)=0$. Moreover, to insure continuity at the origin the
values of the fields at $r=0$ must be $A(0)=1$, $B(0)=0$, $C(0)=0$ and
$G(0)=0$. 
Therefore we have to give up the simplifying choice $C(r)\equiv 0$ and to
include the $C$-field into the formulae. This is, however, no substantial
difficulty.

In substituting the decomposition \ur{dec} into \eq{eig} one has to
calculate matrix elements
in the $\ket{K,J,L;K_3}$ basis of the scalar operators
appearing in \eq{eig}, namely $(\bm\sigma\bf n), (\bm\sigma\bm\tau),
(\bm\tau\bf n), (\bm\sigma\bm\partial),
([\bm\sigma \times \bm\tau]{\bf n})$ and $(\bm\sigma\bf n)(\bm\tau
\bf n)$. We call these operators "scalar" as they commute with the operator
$\op{K}^2$, hence equations for different $K$ will not be mixed.

Thus, the Dirac Hamiltonian \ur{ham} is block-diagonal in the
$\ket{K,J,L;K_3}$ basis, where the blocks correspond to sectors with definite
$K$ and $K_3$; we call these blocks ${\cal H}_{KK_3}$. Since they do not depend
on $K_3$ this index will be suppressed subsequently. Each ${\cal H}_K$ is
a $8 \times 8$ matrix acting on an 8-vector $V_K$:
\beq
{\cal H}_KV_K=EV_K,
\la{eigK}\eeq
where
\begin{eqnarray}
V_K&=&\Bigl(f_{K+\frac{1}{2},K+1},\;\;\;
f_{K+\frac{1}{2},K},\;\;\; f_{K-\frac{1}{2},K},\;\;\;
f_{K-\frac{1}{2},K-1},
\nonumber \\
&&\;\; g_{K+\frac{1}{2},K+1},\;\;\;
g_{K+\frac{1}{2},K},\;\;\; g_{K-\frac{1}{2},K},\;\;\;
g_{K-\frac{1}{2},K-1}\Bigr),
\la{vector} \end{eqnarray}
and ${\cal H}_K$ can be presented as
\beq
{\cal H}_K=\matr{{\cal D}_K+{\cal V}_K}{{\cal W}_K}{{\cal W}^\dagger_K}
{-{\cal D}_K}.
\la{8}\eeq
Here ${\cal D}_K$ is a $4 \times 4$ matrix composed of
differentiation operators:
\beq {\cal D}_K=\left(\begin{array}{cccc} 0 &
-\frac{d}{dx}+\frac{K}{x} & 0 & 0 \\ \frac{d}{dx}+\frac{K+2}{x} & 0 & 0 & 0
\\ 0 & 0 & 0 & -\frac{d}{dx}+\frac{K-1}{x} \\ 0 & 0 &
\frac{d}{dx}+\frac{K+1}{x} & 0 \end{array} \right),
\la{Dif}\eeq
while ${\cal V}_K$ and ${\cal W}_K$ are also $4 \times 4$ matrices made of
the background $W$ and $H$ fields. To shorten notations we introduce
\[
b_K=\sqrt{K(K+1)}, \;\;\,\; c_K=2K+1,
\]
\beq A_K=\frac{1-A(x)}{x c_K},
\;\; B_K=\frac{B(x)}{x c_K}, \;\; C_K=\frac{C(x)}{x c_K}, \;\;
G_K=\frac{G(x)}{c_K},\;\; H_K=H(x).
\la{obozn}\eeq
Using standard $6j$ symbols as described in app.~C, we obtain
for the matrix ${\cal V}_K$:
\beq
\left(\begin{array}{cccc}
B_K(K+1)-\frac{C_K}{2} & -A_K(K+1) & A_Kb_K & (B_K-C_K)b_K \\
-A_K(K+1) & -B_K(K+1)-\frac{C_K}{2} & (B_K+C_K)b_K & -A_Kb_K \\
A_Kb_K & (B_K+C_K)b_K & -B_KK+\frac{C_K}{2} & A_KK \\
(B_K-C_K)b_K & -A_Kb_K & A_KK & B_KK+\frac{C_K}{2}
\end{array}\right)
\la{V}\eeq
and
\beq
{\cal W}_K=m_F\left(\begin{array}{cccc}
H_K & G_K & -2G_Kb_K & 0 \\
-G_K & H_K & 0 & -2G_Kb_K \\
2G_Kb_K & 0 & H_K & -G_K \\
0 & 2G_Kb_K & G_K & H_K
\end{array}\right).
\la{W}\eeq
\indent
In \eqsss{eigK}{W} we imply that $x$ is the distance from the origin
measured in  units of $m_W^{-1}$; the energy eigenvalues $E$ and the fermion
mass $m_F$ are measured in units of $m_W$.

For $K=0$ \eqsss{eigK}{W} reduce to those derived in ref. \cite{KB}; for
any $K$ but $A=B=0$ we get to the equations derived in a different context
in ref. \cite{DPP}.

In order to find the Dirac eigenvalues numerically one
can use various methods. One method \cite{DPP} is to express
fermionic observables (such as the energy of the
Dirac sea, etc.) through the phase shifts of the Hamiltonian ${\cal H}_K$.
In this work we shall use an alternative method: the diagonalization
of the Hamiltonian ${\cal H}_K$ \ur{8} in the so-called Kahana--Ripka basis
\cite{KR}. This method was previously used in the chiral nucleon
model \cite{Bo}. The idea is to discretize the spectrum by putting the fermions
into a large spherical box, and diagonalizing the Hamiltonian in the basis of
spherical Bessel functions which are eigenfunctions of the zero-field
Hamiltonian. We will now introduce this technique.

In the absence of the background field 
we have $A(x)=1, \; B(x)=0, \;  C(x)=0, \;
H(x) = 1$ and $G(x) = 0$, therefore ${\cal V}_K = 0, \; {\cal W}_K = m_F$,
so that the free Hamiltonian becomes
\beq
{\cal H}_K^{(0)}=\matr{{\cal D}_K}{m_F}{m_F}{-{\cal D}_K},
\la{80}\eeq
and the eigenvalue equation can be easily solved analytically.
For given grand spin $K$ and momentum $p$ we find, generally speaking, eight
linearly independent solutions of \eq{eigK}, which we call $v_K^{(q)},
\;\; q=1\ldots 8$; each of them is an 8-vector in the sense of
\eq{vector}. Introducing a column composed of the spherical Bessel
functions,
\begin{eqnarray}
J_{K,i}& =& \Bigl(j_{K+1}(px), \;\; j_K(px),  \;\; j_K(px), \;\; j_{K-1}(px),
\nonumber \\
 &&\ \,j_{K+1}(px), \;\; j_K(px), \;\; j_K(px), \;\;
j_{K-1}(px) \Bigr), \;\;\; i=1\ldots 8,\qquad\quad
\la{Bessel} \end{eqnarray}
the eight solutions of the free eigenvalue equation can be written as
\beq
v_{K,i}^{(r)} = {\cal N} C_i^{(r)}J_{K,i} \;\;\;({\rm no \;\;
summation \;\; in \;\; i \; !})\,,
\la{v}\eeq
where ${\cal N}$ is the normalization factor (see below). With
$\e\equiv\sqrt{p^2+m_F^2}$, the factors $C_i^{(r)}$ are given by
\[
C_i^{(1)}=\frac{1}{\sqrt{2}}
\left(\frac{m_F}{\e},0,0,0,1,-\frac{p}{\e},0,0\right), \;\;\;
C_i^{(5)}=\frac{1}{\sqrt{2}}
\left(-\frac{m_F}{\e},0,0,0,1,\frac{p}{\e},0,0\right),
\]\[
C_i^{(2)}=\frac{1}{\sqrt{2}}
\left(-\frac{p}{\e},-1,0,0,0,-\frac{m_F}{\e},0,0\right), \;\;
C_i^{(6)}=\frac{1}{\sqrt{2}}
\left(\frac{p}{\e},-1,0,0,0,\frac{m_F}{\e},0,0\right),
\]
\[
C_i^{(3)}=\frac{1}{\sqrt{2}}
\left(0,0,\frac{m_F}{\e},0,0,0,1,-\frac{p}{\e}\right), \;\;\;
C_i^{(7)}=\frac{1}{\sqrt{2}}
\left(0,0,-\frac{m_F}{\e},0,0,0,1,\frac{p}{\e}\right),
\]
\beq
C_i^{(4)}=\frac{1}{\sqrt{2}}
\left(0,0,-\frac{p}{\e},-1,0,0,0,-\frac{m_F}{\e}\right), \;\;
C_i^{(8)}=\frac{1}{\sqrt{2}}
\left(0,0,\frac{p}{\e},-1,0,0,0,\frac{m_F}{\e}\right).
\la{c}\eeq
\indent
It is convenient to discretize the free Dirac spectrum by the following
trick \cite{KR,Bo} which preserves the spherical symmetry of the problem and
makes the eight states $v_K^{(r)}$ orthogonal to each other and to any other
states with different $K$ and $\e$. Namely, we introduce a large radius
$R$ (eventually to be taken to infinity) and fix the spectrum for
given $K$ as zeros of the spherical Bessel function:
\beq
j_K(p_nR)=0.
\la{zeros}\eeq
If $\alpha_{m,n}$ are zeros of $j_K(z)$ one has the following
ortho-normalization conditions:
\begin{eqnarray}
&&\quad\int_0^1 dt \,t^2 j_K(\alpha_m t) j_K(\alpha_n t)
\nonumber \\
&&= \int_0^1 dt \,t^2 j_{K\pm 1}(\alpha_m t) j_{K\pm 1}(\alpha_n t) =
\delta_{mn} \frac{1}{2} \left[j_{K\pm 1}(\alpha_n)\right]^2.
\la{ort} \end{eqnarray}
\indent
These relations provide the orthogonality of the eight
degenerate states \ur{v} as well as their orthogonality to states with
different $p_n$ at given $K$. States with different values of $K$ are
orthogonal owing to the angular integration.
{}From \eq{ort} we also learn that the normalization factor ${\cal N}$ in
\eq{v} is
\beq
{\cal N} = \sqrt{\frac{2}{R^3}}\;|j_{K\pm 1}(p_nR)|^{-1}\,.
\la{Norm}\eeq
\indent
We have thus constructed a complete ortho-normalized set of states
$v_K^{(q)}(p_n)$ (given by \eqsss{Bessel}{c}) which are the
eigenfunctions of the free Dirac operator.

The eigenstates of the full Dirac Hamiltonian \ur{8} can be now found
from a direct diagonalization of the matrix
\beq
{\cal H}_K^{rs}(p_m,p_n)= \int_0^R dx\,x^2 v_{K,i}^{(r)}(p_m,x){\cal
H}_{K,ij} v_{K,j}^{(s)}(p_n,x)
\la{diag}\eeq
(summation over $i,j=1...8$ is
assumed here). The superscripts $r,s$ run over $1\ldots 8$ while the radial
momenta $p_{m,n}$ are given by \eq{zeros}. They vary from small values of
the order of $1/R$ up to some numerical cut--off $P_{max}$, which ensures
the finiteness of the basis. $P_{max}$ and $R$ have to be taken large enough
so that no result changes when they are further increased.
 
\section{C}
\setcounter{equation}{0}
\def\theequation{\Alph{section}.\arabic{equation}}

In this appendix we derive matrix elements of scalar operators entering
the Hamiltonians for bosonic and fermionic fluctuations about the spherically
symmetric sphaleron. These operators commute with the grand spin
\beq
\op{K}=\op{J}+\op{T}=\op{L}+\op{S}+\op{T}\,.
\la{gs}
\eeq
The fully normalized set of states in this momentum-adding scheme is
\beq
\ket{L,S,[J],T,K,K_3 }=
\sum_{L_3,S_3,J_3,T_3}
\C{KK_3}{JJ_3}{TT_3} \C{JJ_3}{LL_3}{SS_3}
 \ket{SS_3}\ket{TT_3} \ket{LL_3}\,.
\la{basis}
 \eeq
These states are the eigenfunctions of the operators $\op{K}^2$,
$\opr{K}_3$, $\op{L}^2$, $\op{S}^2$,  $\op{J}^2$ and $\op{T}^2$.
On general grounds one can show that such Hamiltonians depend
on the following operators which commute with the grand spin:
\beq
\op{J}^2 , \;   \op{L}^2 , \;   \op{T}^2 , \;  \op{S}^2 , \;
\op{S}\cdot \op{L}, \;\op{S}\cdot \op{T},
\; \op{T}\cdot \op{n},  \;\op{S}\cdot \op{n},
\; \op{T} \cdot [\op{S} \times \op{n}],
\; \op{T} \cdot [\op{n} \times \op{L}]\,,
\la{opera}
\eeq
where $ \op{n}= \op{x}/|\op{x}|$. We use the general relations
\beq
\partial_k=n_k \frac{\partial}{\partial r} -\frac{i}{r}\varepsilon_{klm}
n_l L_m\, ,\qquad
L_j= -i \varepsilon_{jlm} x_l \partial_m\, .
\eeq
Any scalar operator commuting with the grand spin can be expressed
through the set of operators (\ref{opera}), for example
\beq
\op{T}\cdot \op{L}=\frac12 (\op{K}^2 -\op{J}^2 -\op{T}^2  ) -
\op{S}\cdot \op{T}\quad {\rm or}\quad
\op{S}\cdot \op{L}=\frac12 (\op{J}^2 -\op{S}^2 -\op{L}^2  )\,.\quad
\eeq
To calculate matrix elements of the operators (\ref{opera})
in the basis \ur{basis} we use the technique of irreducible tensor
operators and the Wigner--Eckart theorem (for details see
\cite{Varsh}).

We illustrate this technique by calculating the complicated matrix element
$
\bra{L^\prime,S^\prime ,[J^\prime ],T^\prime ,K,K_3 }
\,\op{T} \cdot (\op{S} \times \op{n}) \,
\ket{L,S,[J],T,K,K_3 }
$.
The operator $\op{T}$ does not act on the spin and angular variables
so that the matrix element can be factorized as
\begin{eqnarray}
&&\bra{L^\prime,S^\prime ,[J^\prime ],T^\prime ,K,K_3 }\;
 \op{T} \cdot (\op{S} \times \op{n})\;
\ket{L,S,[J],T,K,K_3 } \nonumber \\
&& =(-1)^{J+T^\prime+K} \Sj{K}{T^\prime}{J^\prime}{1}{J}{T}
\irre{L^\prime S^\prime [J^\prime] }{\op{S}\times \op{n}}{L S [J] }
\;\irre{T^\prime}{\op{T}}{T}\, ,
\nonumber\\ \la{tsn}\end{eqnarray}
where
\beq
\irre{T^\prime}{\op{T}}{T} = \sqrt{T(T+1)(2T+1)} \;\delta_{TT^\prime}\, .
\eeq
The spin operator $ \op{S}$ acts only on spin variables whereas $\op{n}$
acts on angular variables, so one can again factorize the matrix element
as
\[
\irre{L^\prime S^\prime [J^\prime] }{\op{S}\times \op{n}}{L S [J] }
= -i\sqrt{2} \irre{L^\prime S^\prime [J^\prime] }
{[S^{(1)}\otimes n^{(1)} ]^{(1)}}{L S [J] }
\]
\beq
=  i \sqrt{6(2J+1)(2J^\prime+1)}
   \Nj{L^\prime }{L}{1}{S^\prime}{S}{1}{J^\prime }{J}{1}
    \irre{L^\prime }{\op{n}}{L} \,\irre{S^\prime }{\op{S}}{S}\, .
\eeq
Using the relation
for the $9j$ symbol \cite{Varsh}
\[
   \Nj{L^\prime }{L}{1}{S}{S}{1}{J^\prime }{J}{1}
=
\frac{(J^\prime-L^\prime)(J^\prime+L^\prime+1)-
(J-L)(J+L+1) }{\sqrt{24 S(S+1)(2S+1)}}
\]
\beq
\qquad\cdot
(-1)^{J+L^\prime+S+1}
\Sj{J^\prime }{J}{1}{L }{L^\prime }{S}\,,
\eeq
one eventually finds
\begin{eqnarray}
&&\bra{L^\prime,S^\prime ,[J^\prime ],T^\prime ,K,K_3 }\;
 \op{T} \cdot (\op{S} \times \op{n})\;
\ket{L,S,[J],T,K,K_3 } \nonumber \\
&&\ = \frac{i}{2} (-1)^{2J+2L^\prime+T+S+1+K}
\,[(J^\prime-L^\prime)(J^\prime+L^\prime+1)-
(J-L)(J+L+1) ] \nonumber \\
&&\qquad\cdot
\sqrt{T(T+1)(2T+1)(2J+1)(2J^\prime+1)(2L+1)(2L^\prime+1) } \nonumber \\
&&\qquad\cdot
\Sj{J^\prime }{J}{1}{L }{L^\prime }{S}
\cdot
\Sj{K}{t}{J^\prime}{1}{J}{T}
\cdot
  \Tj{L^\prime }{1}{L }{0}{0}{0}\,.
\end{eqnarray}
Along the same lines one can express matrix elements of operators
of (\ref{opera}) in the basis (\ref{basis}) through $6j$ symbols;
the result is:
\begin{eqnarray}
&&\bra{L^\prime,S^\prime ,[J^\prime ],T^\prime ,K,K_3 }\;
 \op{S} \cdot \op{n}\;
\ket{L,S,[J],T,K,K_3 }\nonumber\\
&&\quad =
(-1)^{J+L+L^\prime+S }\delta_{JJ^\prime }
\sqrt{(2L+1)(2L^\prime+1)S(S+1)(2S+1)} \nonumber\\
&&\qquad\cdot
\Sj{J}{S}{L^\prime}{1 }{L}{S} \cdot
  \Tj{L^\prime }{1}{L }{0}{0}{0},\\
\nonumber
\end{eqnarray}
\begin{eqnarray}
&&\bra{L^\prime,S^\prime ,[J^\prime ],T^\prime ,K,K_3 }\;
 \op{T} \cdot \op{n}\;
\ket{L,S,[J],T,K,K_3 }\nonumber\\
&&\quad=
(-1)^{2J+2L^\prime+S+T+K+1 } \nonumber\\
&&\qquad\cdot
\sqrt{(2L+1)(2L^\prime+1)(2J+1)(2J^\prime+1) T(T+1)(2T+1)}\nonumber\\
&&\qquad\cdot
\delta_{SS^\prime } \delta_{TT^\prime }
\Sj{K}{T}{J^\prime}{1 }{J}{T} \cdot
\Sj{L^\prime}{J^\prime}{S}{J }{L}{1} \cdot
  \Tj{L^\prime }{1}{L }{0}{0}{0},\qquad\\
\nonumber
\end{eqnarray}
\begin{eqnarray}
&&\bra{L^\prime,S^\prime ,[J^\prime ],T^\prime ,K,K_3 }\;
 \op{S} \cdot \op{T} \;
\ket{L,S,[J],T,K,K_3 }\nonumber \\
&&\quad=
(-1)^{J+J^\prime+L+S+T+K+1}\nonumber\\
&&\qquad\cdot
\sqrt{(2J+1)(2J^\prime+1)(2S+1)S(S+1) T(T+1)(2T+1)} \nonumber \\
&&\qquad\cdot\delta_{SS^\prime } \delta_{TT^\prime }\delta_{LL^\prime }
\Sj{K}{T}{J^\prime}{1 }{J}{T} \cdot
\Sj{S}{J^\prime}{L}{J}{S }{1}, \\
\nonumber
\end{eqnarray}
\begin{eqnarray}
&&\bra{L^\prime,S^\prime ,[J^\prime ],T^\prime ,K,K_3 }\;
 \op{T} \cdot (\op{n} \times \op{L}) \;
\ket{L,S,[J],T,K,K_3 }\nonumber \\
&&\quad=\frac{1}{2i}(-1)^{J+L^\prime+L+T+S+K}\,
[L^\prime(L^\prime+1)-L(L+1) ] \nonumber \\
&&\qquad\cdot\sqrt{T(T+1)(2T+1)(2J+1)(2J^\prime+1)(2L+1)(2L^\prime+1) }
\nonumber \\
&&\qquad\cdot\Sj{J^\prime }{J}{1}{L }{L^\prime }{S}
\cdot
\Sj{K}{T}{J^\prime}{1}{J}{T}
\cdot
  \Tj{L^\prime }{1}{L }{0}{0}{0}\,.
\end{eqnarray}
We checked these expressions using the commutation relations:
\begin{eqnarray}
\left[\op{S}\cdot \op{T},\op{S}\cdot \op{n}\right]_{-} 
&=&-i\,\op{T} \cdot (\op{S} \times\op{n})\,, \\
\left[\op{S}\cdot \op{T},\op{T}\cdot \op{n}\right]_{-} 
&=& i\,\op{T} \cdot (\op{S} \times\op{n})\,,\\
\left[\op{S}\cdot \op{n},\op{T}\cdot \op{n}\right]_{-} &=& 0\,.
\end{eqnarray}
For practical calculations the $6j$ symbols and Clebsh-Gordan coefficients
can be evaluated using e.g.~{\it Mathematica}.
 
\section{D}
\setcounter{equation}{0}
\def\theequation{\Alph{section}.\arabic{equation}}

In this appendix we state the divergent parts of the sea energy
\ur{sqr} in the basis of the set of eigenfunctions of the free
Dirac-Hamiltonian, given by \eqsss{Bessel}{c}. To this end a
semiclassical expansion up to the quadratically divergent terms is
performed. Subtraction of the result
from the sea energy removes its quadratically divergent part for
each value of the grand spin $K$ separately. Since the quadratically
divergent term is already complicated and lengthy in this basis,
we do not continue the expansion to include also the logarithmic
divergencies. Instead we use a simpler distribution of the total
logarithmic divergence among the $K$ sectors
which may leave the separate sectors logarithmically divergent, but
renders the sum over all $K$ finite. Since the total value of the
logarithmic divergence is small, the complete removal of all divergencies
for each $K$ is of little use and would cause an enormous increase
of numerical effort.

We start from \eq{exactK} , insert \eqs{v}{c}, perform the sum
$\sum_r$ and obtain:
\begin{eqnarray}
&&E_{div}=\sum_{K=0}^\infty(2K+1)\frac{1}{4\sqrt{\pi}}
\int_0^1\frac{dt}{t^{3/2}}
\sum_{n=1}^\infty \frac{2e^{-tp_n^2}}{j_{K+1}^2(p_nR)R^3}
\nonumber\\
&&\qquad\quad\cdot\int_0^R dx\,x^2\sum_{i=1}^8 J_{K,i}(p_nx)
\left[e^{-t({\cal H}_K^2-p_n^2)}\right]_{ii}J_{K,i}(p_nx)
\Biggr|_{div}\, .
\la{EdivK} \end{eqnarray}
We use the expansion
\beq
e^{-t({\cal H}_K^2-p_n^2)}=1-t({\cal H}_K^2-p_n^2)
+\frac{t^2}{2}({\cal H}_K^2-p_n^2)^2+\ldots\ .
\la{Hexp}\eeq
The constant $1$ is cancelled by subtraction of the vacuum (free field).
Contributions to the quadratic divergence arise from the terms ${\cal O}(t)$,
${\cal O}(t^2p_n^2)$ and ${\cal O}(t^2K^2)$. The result is as follows:
\[
E_{div,2}=\sum_{K=0}^\infty(2K+1)\frac{1}{2\sqrt{\pi}}
\int_0^1\frac{dt}{t^{1/2}}
\sum_{n=1}^\infty \frac{e^{-tp_n^2}}{j_{K+1}^2(p_nR)R^3}\int_0^R dx\,
x^2 F(x),
\]
with
\[
F(x)=-\left\{\delta_{K,0}\left[2m_F^2(j_1^2+j_0^2)(H^2+G^2-1)
   + (j_1^2+j_0^2)\left(\At^2+\Bt^2+\frac{\Ct^2}{4}\right)\right.\right.
\]\[
\left.+ (j_1^2-j_0^2)(\At'-\Bt\Ct)+(-3j_1^2-j_0^2)
\left(\frac{\At}{x}\right)\right]
\]\[
+(1-\delta_{K,0})\Biggl[2m_F^2(j_{K+1}^2+2j_K^2+j_{K-1}^2)(H^2+G^2-1).
\]\[
+\left(\frac{K+1}{2K+1}j_{K+1}^2+j_K^2+\frac{K}{2K+1}j_{K-1}^2\right)
(\At^2+\Bt^2)
+(j_{K+1}^2+2j_K^2+j_{K-1}^2)\left(\frac{\Ct^2}{4}\right)
\]\[
+\left(\frac{K+1}{2K+1}j_{K+1}^2-j_K^2+\frac{K}{2K+1}j_{K-1}^2\right)
(\At'-\Bt\Ct)
\]\[
\left.+\left(-\frac{(K+1)(2K+3)}{2K+1}j_{K+1}^2-j_K^2
+\frac{K(2K-1)}{2K+1}j_{K-1}^2\right)\left(\frac{\At}{x}\right)
\Biggr]\right\}
\]\[
+\frac{t}{2}\left\{\delta_{K,0}\left[(9j_1^2+j_0^2)\left(\frac{\At}{x}
\right)^2 + (j_1^2+j_0^2)\left(\frac{\Bt}{x}\right)^2\right.\right.
\]\[
\left.+(j_1^2+j_0^2)\Ct^2p_n^2 +\left(-\frac{9}{4}j_1^2
-\frac{1}{4}j_0^2\right)\left(\frac{\Ct}{x}\right)^2\right]
\]\[
+(1-\delta_{K,0})\Biggl[\left(\frac{(K+1)(2K^2+7K+9)}{2K+1}j_{K+1}^2
+(2K^2+2K+1)j_K^2\right.
\]\[
\left.+\frac{K(2K^2-3K+4)}{2K+1}j_{K-1}^2\right)
\left(\frac{\At}{x}\right)^2
\]\[
+\left((K+1)^2j_{K+1}^2+(2K^2+2K+1)j_K^2+K^2j_{K-1}^2\right)
\left(\frac{\Bt}{x}\right)^2
\]\[
+(j_{K+1}^2+2j_K^2+j_{K-1}^2)\Ct^2p_n^2
\]\[
+\left(-\frac{8K^3+44K^2+50K+9}{4(2K+1)}j_{K+1}^2
-\frac{4K^2+4K+3}{2}j_K^2\right.
\]\beq
\left.\left. -\frac{8K^3-20K^2-14K+5}{4(2K+1)}j_{K-1}^2
\right)\left(\frac{\Ct}{x}\right)^2\Biggr]\right\},
\la{Fx}\eeq
where we have omitted the argument $(p_nx)$ of the Bessel-functions
and denoted:
\beq
\At\equiv\frac{1-A(x)}{x},\,\,\,\,\,
\Bt\equiv\frac{B(x)}{x},\,\,\,\,\,
\Ct\equiv\frac{C(x)}{x},\,\,\,\,\,
\At'\equiv-\frac{1}{x}\frac{dA}{dx}(x).
\la{til}\eeq
\indent
To see whether this expression corresponds to the quadratically
divergent part of \eq{semicl} we perform the limit $R\to\infty$
in which
$j_{K+1}^2(p_n)R^3\to R/p_n^2$ and $1/R\sum_n\to 1/\pi\int dp$.
Using
\beq
\sum_K (2K+1)j_K^2(px)=1, \,\,\,\,\,\,\,\,\,\, \sum_K (2K+1)^3j_K^2(px)
=1+\frac{8}{3}(px)^2,
\la{suK}\eeq
and performing the $p$ integration we obtain:
\[
E_{div,2}=-\frac{1}{\pi}\int_0^1\frac{dt}{t^2}
\int_0^\infty dx\, x^2m_F^2(H^2+G^2-1)\qquad
\]\beq
\qquad\quad+\frac{1}{4\pi}\int_0^1\frac{dt}{t}
\int_0^\infty dx\left(\At^2+\Bt^2+\frac{5\Ct^2}{4}\right).
\la{Rlim}\eeq
The first term coincides with the quadratically divergent part
of \eq{semicl}, as it should do. The second term is logarithmically
divergent so that
the remaining logarithmic divergence of \eq{semicl} to be removed is
therefore:
\[
E_{div,1}=\frac{1}{2\pi}\int_0^1\frac{dt}{t}\int_0^\infty dx\, g(x),
\]
\begin{eqnarray}
&&g(x)=\frac{1}{6}\left[\left(A'+\frac{BC}{x}\right)^2 +
    \left(B'-\frac{AC}{x}\right)^2 +\frac{(A^2+B^2-1)^2}{2x^2}\right]
\nonumber \\
 &&\quad\qquad + m_F^4 x^2\left[(G^2+H^2)^2-1\right]
\nonumber \\
&&\quad\qquad+ \frac{m_F^2}{2}\left[(G^2+H^2)
 (1+A^2+B^2+C^2/2)+ 2A(G^2-H^2)-4BGH\right.
\nonumber \\
&&\quad\qquad\left.+ 2x^2({G'}^2+{H'}^2) - 2xC(HG'-H'G)\right]
  -\frac{(1-A)^2+B^2+\frac{5C^2}{4}}{2x^2}\; . \nonumber \\
\la{ldiv}\end{eqnarray}
This divergence can be distributed among the $K$ sectors as follows:
\[
E_{div,1}=\sum_{K=0}^\infty(2K+1)\frac{1}{2\sqrt{\pi}}
\int_0^1 dt\,t^{1/2}
\sum_{n=1}^\infty \frac{e^{-tp_n^2}}{j_{K+1}^2(p_nR)R^3}\int_0^R dx\,
x^2 G(x)
\]
with
\begin{eqnarray}
&&G(x)=\left[\delta_{K,0}\left(j_1^2(p_nx)+j_0^2(p_nx)\right)\right.
\nonumber \\
 &&\qquad\qquad \left.+ (1-\delta_{K,0})\left(j_{K+1}^2(p_nx)+2j_K^2(p_nx)
+j_{K-1}^2(p_nx)\right)\right]\frac{g(x)}{x^2}\;.\qquad\quad
\la{logK}\end{eqnarray}
As already mentioned, we would obtain a much more complicated distribution
of the logarithmic divergence among the $K$ sectors if we continued the
expansion \eq{Hexp} up to fourth order and collected the logarithmically
divergent terms. Although the simpler version \eq{logK} removes the
logarithmic divergencies only after summing over $K$ and not for each
$K$ separately, it is sufficient for all numerical purposes to use this
formula.

Let us finally remark that one can in principle use a simpler
formula also for the quadratically divergent terms which can be obtained
by expanding \eq{Hexp} in a large $K$ limit. This leads to an expression for
$E_{div,2}$ similar to \eq{Fx} but contains only the first term
$H^2+G^2-1$. Numerically identical results are obtained with this version,
but one needs higher numerical parameters to insure stability, 
so that the numerical effort is increased.
\newpage\noindent
{\large\bf Figure captions}
\begin{enumerate}
\item Energy barrier as a function of $N_{CS}$ for different values of the
chemical potential $\mu$ for $m_H = m_W$. The units for $E_{class}(\mu)$ 
and $\mu$ are $M_0=\mu_{crit}=2\pi m_W/\alpha$.
\item The discrete level $\e_{val}$ and some discretized continuum eigenstates
as a function of $N_{CS}$ for $m_F=m_W$. It is demonstrated
that each level is shifted upwards and finally takes the position
of its predecessor.
\item The renormalized fermionic energy $E_{fer}^{ren}$ as a function
of $N_{CS}$ for the fermion masses
$m_F/m_W=0$, $1$ and $2$. It has to be compared with the classical
energy which is about $100\, m_W$ for $N_{CS}=0.5\,$.
\item The fermionic temperature dependent part $\beta E_{fer}^{temp}$
(solid lines)
and the classical part $\beta E_{class}^{ren}$ (dashed lines)
of the sphaleron transition rate as a function of $T$ for
$m_t/m_W=1.5$, $2.0$ and $2.5$. $E_{class}^{ren}$ and $E_{fer}^{temp}$
are defined in \eqs{transsplit}{transterms}.
\item The baryon density as a function of the radial distance $r$ for
$m_F=m_W$ and $N_{CS}=0.26$, $0.5$, $0.74$, and close to $1$.
\end{enumerate}
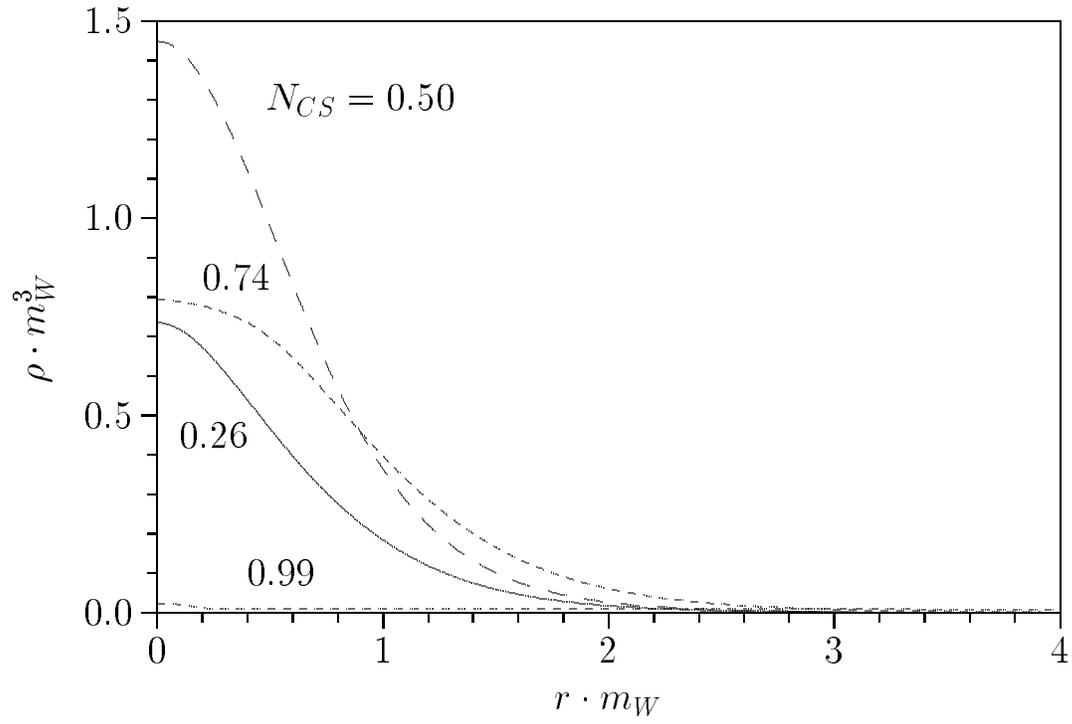
\end{document}